\documentclass[aps,prd,onecolumn,nofootinbib,amsmath,amssymb,floatfix,letterpaper,superscriptaddress,notitlepage]{revtex4-1}
\usepackage{graphicx}
\usepackage{hyperref}
\usepackage{color}
\usepackage{amsmath,amssymb}
\usepackage{bm}
\newcommand{\beq}{\begin{equation}}
\newcommand{\eeq}{\end{equation}}
\newcommand{\bea}{\begin{eqnarray}}
\newcommand{\eea}{\end{eqnarray}}
\newcommand{\bsp}{\begin{split}}
\newcommand{\esp}{\end{split}}

\newcommand{\hMpc}{\ h^{-1}\text{Mpc}}

\newcommand{\vq}{\vec q}
\newcommand{\confh}{\mathcal{H}}
\newcommand{\ihMpc}{\; h\text{Mpc}^{-1}}
\newcommand{\hMpcsq}{\; h^{-2}\text{Mpc}^{2}}
\newcommand{\cs}{c_\text{s}^2}
\newcommand{\derd}{\text{d}}
\newcommand{\ddir}{\delta^{\text{(D)}}}
\newcommand{\la}{\left\langle}
\newcommand{\ra}{\right\rangle}

\renewcommand{\vec}[1]{\bm{#1}}

\definecolor{darkgreen}{RGB}{0,120,0}

\begin{document}
\title{The Effective Field Theory of Large Scale Structure at Two Loops: the apparent scale dependence of the speed of sound}
\author{Tobias Baldauf}
\email{baldauf@ias.edu}
\affiliation{Institute for Advanced Study, Princeton, NJ, USA}
\author{Lorenzo Mercolli}
\affiliation{Department of Astrophysical Sciences, Princeton University, Princeton, NJ, USA}
\affiliation{Federal Office of Public Health FOPH, Bern, Switzerland}
\author{Matias Zaldarriaga}
\affiliation{Institute for Advanced Study, Princeton, NJ, USA}
\begin{abstract}
We study the Effective Field Theory of Large Scale Structure for cosmic density and momentum fields. 
We show that the finite part of the two-loop calculation and its counterterms introduce an apparent scale dependence for the leading order parameter $c_\text{s}^2$ of the EFT starting at $k=0.1\ihMpc$. These terms limit the range over which one can trust the one-loop EFT calculation at the $1\%$ level to $k<0.1\ihMpc$ at redshift $z=0$.  
We construct a well motivated one parameter ansatz to fix the relative size of the one- and two-loop counterterms using their high-$k$ sensitivity.  Although this one parameter model is a very restrictive choice for the counterterms, it explains the apparent scale dependence of $c_\text{s}^2$ seen in simulations. It is also able to capture the scale dependence of the density power spectrum up to $k\approx 0.3\ihMpc$ at the $1\%$ level at redshift $z=0$. Considering a simple scheme for the resummation of large scale motions, we find that the two loop calculation reduces the need for this IR-resummation at $k<0.2\ihMpc$. Finally, we extend our calculation to momentum statistics and show that the same one parameter model can also describe density-momentum and momentum-momentum statistics.
\end{abstract}
\maketitle
\section{Introduction}
The  development of an Effective Theory for Large Scale Structure (EFT of LSS) \cite{Baumann:2010tm,Carrasco:2012cv}
has lead to a resurgence of interest in perturbative approaches to study the development of structure in our Universe. Although Standard Perturbation Theory (SPT, see e.g.~\cite{Bernardeau:2001qr}) has allowed invaluable insights in the physics of LSS, it was soon realized that in order to extend the validity of the theory down to smaller scales, one needs do go beyond SPT. Various approaches that can be found in the literature, e.g.~\cite{Crocce:2005xy,Bernardeau2008,Taruya2012,Blas:2013aba}, focus on the resummation of higher order contributions in order to achieve an accurate description of non-linear data up to large wavenumbers. Perturbation theory will never be able to capture the small scale dynamics even after complete resummation (as an illustration see \cite{McQuinn:2015tva}).  This fact limits the applicability of resummation results. The EFT of LSS aims at extending SPT through the modeling the effects of small scale dynamics on larger scales. 
Based on the same principles that high-energy physics community has been exploiting for decades, the EFT framework allows to describe perturbatively the evolution of long wavelength modes while systematically taking into account the impact that short wavelength modes can have. The power of this approach lies in the fact that it is not necessary to explicitly solve the non-linear small-scale dynamics, which, however, comes at the cost of introducing parameters that are not determined by the theory itself. Furthermore, the EFT approach allows to overcome some conceptual shortcomings of SPT.

Since the original papers \cite{Baumann:2010tm,Carrasco:2012cv}, many aspects of the EFT of LSS have been explored in the literature. At the one-loop level \cite{Hertzberg2014,Pajer:2013jj,Mercolli:2013bsa,Carroll2014,Senatore:2014via,Senatore2014a,Foreman:2015uva} have made progress, while in \cite{Carrasco:2013sva,Carrasco:2013mua} attempts were made to tackle the EFT of LSS at the two-loop order. These references mainly focussed on the two-point functions, while \cite{Baldauf:2015qfa, Angulo:2014tfa} considered the matter bispectrum and \cite{Angulo2015,Assassi2015} looked at non-Gaussian effects. The Lagrangian space formulation of the EFT of LSS has been studied in \cite{Porto:2013qua,Vlah2015} and finally, aspects of bias and baryonic effects have been considered in \cite{McDonald2006,McDonald2009,Schmidt2013,Assassi2014a,Senatore2014,Mirbabayi2014,Lewandowski2015,Angulo2015}.

Despite the power of the EFT approach, we have to deal with the presence of free parameters. For the power spectrum at the one-loop level, one such parameter is introduced. For the bispectrum, three additional parameters are necessary and for the two-loop power spectrum an even larger number of free parameters would have to be considered.\footnote{Consider e.g.~Chiral Perturbation Theory where in the strong sector at the leading order there are two, at the next higher order 12 and at the third order more than 100 free parameters.} This, however, means that the theoretical description stops being predictive and three- and four-point functions would have to be considered in order to determine the values of all parameters through a comparison with simulations or observations.

The role of the free parameters and the corresponding counterterms is to incorporate the effects of the small scales into the theory. Our aim is to achieve exactly this in a systematic way for the power spectrum at the two-loop level. The approach that we follow does, however, avoid an unmanageable number of free parameters by making a well motivated ansatz. 

In this paper we revisit the two-loop Eulerian power spectrum calculation  and compare the results to our own set of numerical simulations. We compare results at the level of the power spectra as was done in \cite{Carrasco:2013sva,Carrasco:2013mua}. In two companion papers \cite{Baldauf:2015tla, Baldauf:2015tlb} we compare perturbation theory with the results of numerical simulations for the same initial conditions. This is a more stringent test than what is presented here. Our goal in this paper is to reproduce the comparison method used in the literature and try to relate the result to what we see in the more detailed comparison. We will find that in terms of the maximum $k$ where the perturbative calculation can be trusted both results agree.

This paper is organized as follows. After a brief review of the EFT of LSS, we consider in more detail the UV sensitivity of the one- and two-loop integrals in Sec.~\ref{sec:uv1} and \ref{sec:uv2}. From the UV sensitivity, we derive our ansatz for the two-loop counterterms in Sec.~\ref{sec:ansatzct} and an even simpler procedure for the counterterms is discussed in Sec.~\ref{sec:simple}. In Sec.~\ref{sec:sims} we compare our approach with numerical simulations and present our results. Also, we discuss the two-point correlations functions that involve momentum.

\section{The EFT of LSS}

In the EFT of LSS one sets to solve perturbatively the following equations:
\bea
\partial_\tau \delta + \partial_i[(1+\delta)v^i]&=&\partial_i u^i  \;, \nonumber  \\
\partial_\tau v^i + \confh v^i + \partial^i \phi +  v^j \partial_j v^i &=& - {1 \over a \rho} \partial_j \tau^{i j}   \;, \label{eqspt}\\
\triangle \phi &=& {3 \over 2} \confh^2 \Omega_\text{m} \delta \;.\nonumber
\eea
These equations differ from those of SPT \cite{Bernardeau:2001qr} due to the addition of new source term, $u^i$ in the continuity equation and a stress tensor source $\tau^{ij}$ in the Euler equation. These sources arise from small scales, where the perturbative solution of SPT is not applicable. In the EFT of LSS they have to be modeled as they arise from modes that are outside the range of applicability of the theory and thus result in the introduction of free parameters. The EFT of LSS provides an organizing framework for how to model these sources, providing a list of terms with their associated free parameters that need to be introduced to achieve a desired accuracy. 

For simplicity, in the discussion that follows we concentrate on the stresses that appear in the Euler equation. In this paper we will not consider velocity statistics, but only statistics involving the density and the momentum. In such case it suffices to discuss the stresses in the Euler equation as the effects from $u^i$ in the statistics we will consider can be mimicked by changing $\tau^{i j}$. In any case, all the conceptual points we will make below are applicable to both $\tau^{i j}$ and $u^i$. 

The $\tau^{i j}$ stresses come in two different forms. Some of these stresses can be computed in terms of the perturbative solution, others cannot. For the latter one only has a model for the statistical properties of those stresses. It is convenient to decompose the velocity field into its gradient and curl pieces. At the order we will work in this paper only the gradient component will be relevant, thus the stresses we need to model only enter through a scalar quantity:
\beq
\tau_\theta \equiv - \partial_i \biggl[{1 \over a \rho} \partial_j \tau^j\biggr]= \tau_\theta^\text{det} +   \tau_\theta^\text{stoch}. 
\eeq
The deterministic part of the stresses $\tau_\theta^\text{det}$ can be modelled perturbatively. In the EFT we write schematically 
\beq
\tau_\theta^\text{det} = \tau_\theta^\text{det}[\partial_i \partial_j \bar\phi].
\eeq
The deterministic part of the stresses is a local function of the perturbative solution, and we have used the equivalence principle to demand that it can only depend on second derivatives of the gravitational potential (higher spatial derivatives and time derivatives can also appear). We have introduced $\bar\phi = \phi /(3/2 \confh^2 \Omega_m)$ so that $\partial_i \partial_j \bar\phi$ is dimensionless and $\triangle\bar\phi = \delta$. 
For the stochastic part, all we can do is model the statistical properties of $\tau_\theta^\text{stoch}$.

In the EFT of LSS $\tau_\theta^\text{det}$ is modeled as a power series in $\partial_i \partial_j \bar{ \phi}$ and its spatial and time derivatives. In addition to the equivalence principle, mass and momentum conservation restrict the form of both  $\tau_\theta^\text{det}$ and of the statistical properties of $\tau_\theta^\text{stoch}$. In particular in Fourier space $\tau_\theta^\text{det}(k)$ needs to go to zero at least as $k^2$ faster than the density when $k\rightarrow 0$ and the power spectrum of  $\tau_\theta^\text{stoch}$ should go to zero at least as $k^4$.

To calculate the one-loop power spectra in $\Lambda$CDM, only the lowest order piece of $\tau_\theta^\text{det}$ is relevant. It is given by
\beq
\tau_\theta^\text{det}\big|_\text{LO} = - d^2 \triangle \delta_{(1)} =  - d^2 \triangle \triangle \bar\phi_{(1)} \;, \label{eq:tauLO}
\eeq
where $\delta_{(1)}$ is the linear solution of perturbation theory. In this formulation, because $\tau_\theta^\text{det}$ acts as a source in the equations of motion, the time dependence of  $d^2$ will affect the results. In particular it will be relevant to determine the relative sizes of the corrections in the different two point functions involving $\delta$ and $\theta$. 

The case of the one loop bispectrum has already been considered in the literature \cite{Angulo:2014tfa,Baldauf:2015qfa}. In that case the second order counterterms are needed. This introduces three additional parameters for the spatial structure of $\tau_\theta^\text{det}$. One can write:
\beq\label{nlocounter}
\tau_\theta^\text{det}\big|_\text{NLO} = - d^2 \triangle [\delta_{(1)}+\delta_{(2)}] - e_1  \triangle \delta_{(1)}^2 - e_2 \triangle (s_{ij(1)}s^{ij}_{(1)} )- e_3 \partial_i s^{ij}_{(1)} \partial_j  \delta_{(1)},
\eeq
with 
\beq
s_{ij} = \left(\partial_i\partial_j - \frac{1}{3} \delta_{ij}^\text{(K)} \triangle\right) \bar\phi.
\eeq
In principle, $d, e_1, e_2$ and $e_3$  could be fixed by measuring both the power spectra and bispectrum. In practice however, with current simulations there are significant degeneracies among these different parameters. Making an ansatz for the ratios, scaling all counterterms by the same amplitude and fitting for this overall amplitude parameter, seems good enough to explain simulation measurements \cite{Baldauf:2015qfa}. 

In this paper we are interested in performing a two-loop calculation for the power spectrum and thus we would have to model the stresses up to third order in the fields. Modeling these terms will increase the number of parameters even further.  At the level of the two point function however, some of these parameters will be degenerate.  In principle, one could disentangle all the new parameters comparing the predictions with the four point function measured from simulations. In practice the necessary signal to noise ratio to do this is probably not available in the current generation of simulations and a simple ansatz for the ratios of amplitudes of the various terms could be good enough. In any case, in this paper we will only compare results against measurements of the two point function and thus we will not have enough information to distinguish all the parameters. Furthermore, in this type of exercise one runs the risk of overfitting the power spectra simply because one is introducing too many additional free parameters. In order to avoid this, one should compare the results of perturbative calculations with simulations at the level of the fields as was done in \cite{Baldauf:2015tla} for the Lagrangian displacement and in  \cite{Baldauf:2015tlb} for the density.  In this paper we will adopt a simple ansatz for the size of the various counterterms and only keep one overall free amplitude as a parameter. We will discuss this in the next sections.

\subsection{Perturbative solution and counterterms}

\begin{figure}
\centering
\includegraphics[width=0.69\textwidth]{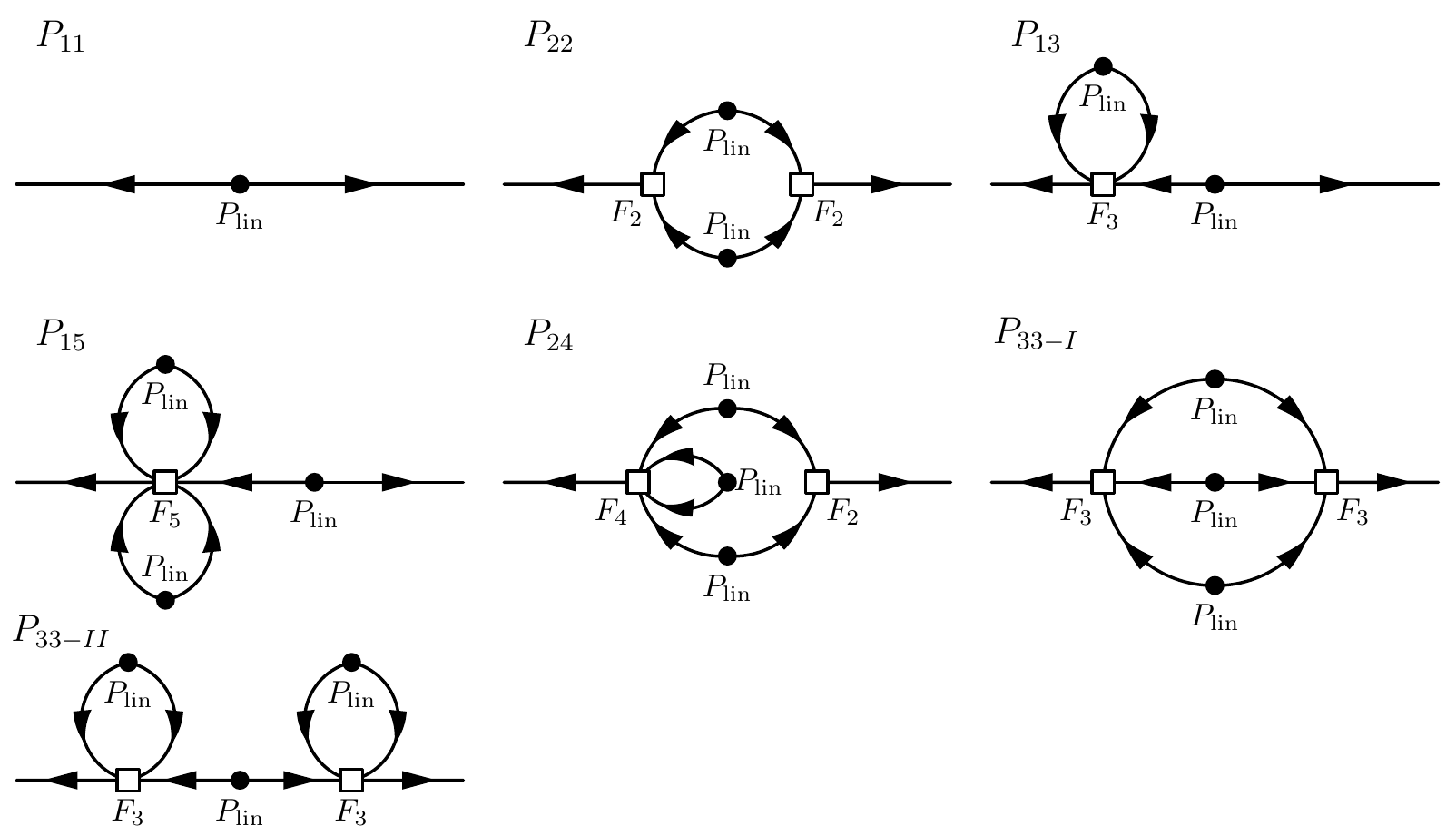}
\caption{Diagrams for the tree level, one- and two-loop expressions of the SPT power spectrum.}
\label{fig:loopdiag}
\end{figure}

In Standard Perturbation Theory (SPT, for a review see\cite{Bernardeau:2001qr}) the perturbative solution of the equations of motion has the following structure, 
\beq
\delta = \delta_{(1)}+\delta_{(2)} + \delta_{(3)}+\delta_{(4)}+\delta_{(5)} +  \cdots
\eeq
where $\delta_{(n)}$ depends on the initial conditions to the $n$-th power and we have only written terms relevant for the two loop calculation of the two point function. When computing the power spectrum, one considers the averages of $\langle \delta_{(n)}\delta_{(m)}\rangle$. At the tree level, the only possible order is $n+m=2$ and therefore $n=m=1$. For one loop, $n+m=4$ so the two possible terms are the mixed term between $ \delta_{(3)}$ and  $\delta_{(1)}$ or the square of $ \delta_{(2)}$. At two loops we have $n+m=6$ so the options are 1-5, 2-4 or 3-3. In SPT one writes the perturbative solutions as\footnote{Note our shorthand notation for the integral and measure \[ \int_{\vq}\equiv \int \frac{\derd^3 q}{(2\pi)^3}\; . \] Furthermore we will express momenta with respect to the external momentum as $q_2=r_2 k$ and $q_1=r_1 k$.
}
\beq
\delta_{(n)}(\vec k) = \int_{\vec q_1} \ldots \int_{\vec q_n} (2\pi)^3\delta^\text{(D)}(\vec q_1 + \ldots \vec q_n-\vec k) F_n(\vec q_1, \ldots, \vec q_n)  \delta_0(\vec q_1) \ldots \delta_0(\vec q_n),
\eeq
where $ \delta_0$ stands for the initial density fluctuations. 
The different contributions to the power spectrum computation can be represented using the diagrams  in Fig.~\ref{fig:loopdiag} and combine to the power spectrum as
\beq
P_{\delta\delta}=P_{11}+2 P_{13}+P_{22}+2 P_{15}+2 P_{24}+P_{33\text{-I}}+P_{33\text{-II}}\; .
\eeq
The explicit expressions for the constituent power spectra are given in App.~\ref{app:expressions}. The integrals for the one- and two-loop contributions to the above expression bear some UV-sensitivity or can be even divergent for certain input power spectra. The EFT provides a framework in which these UV-sensitivities can be addressed and regularized with the corresponding counterterms. That is to say, that the EFT counterterms provided by the stress tensor and its time dependence in Eq.~\eqref{eqspt} should be able to capture and correct the UV-sensitivity of the SPT expression.

The equations of motion (\ref{eqspt}) only have quadratic non-linearities, so vertices in diagrams should only be cubic. That is to say, the $F_n$ kernels in the diagrams we showed in Fig.~\ref{fig:loopdiag} are effective time integrated diagrams that can be constructed by having multiple cubic vertices joined by propagators (or Green's functions) \cite{Crocce:2005xy}. In the EFT, there are additional diagrams due to the introduction of counterterms, or sources in the equations of motion.
\begin{figure}[t]
\centering
\includegraphics[width=0.69\textwidth]{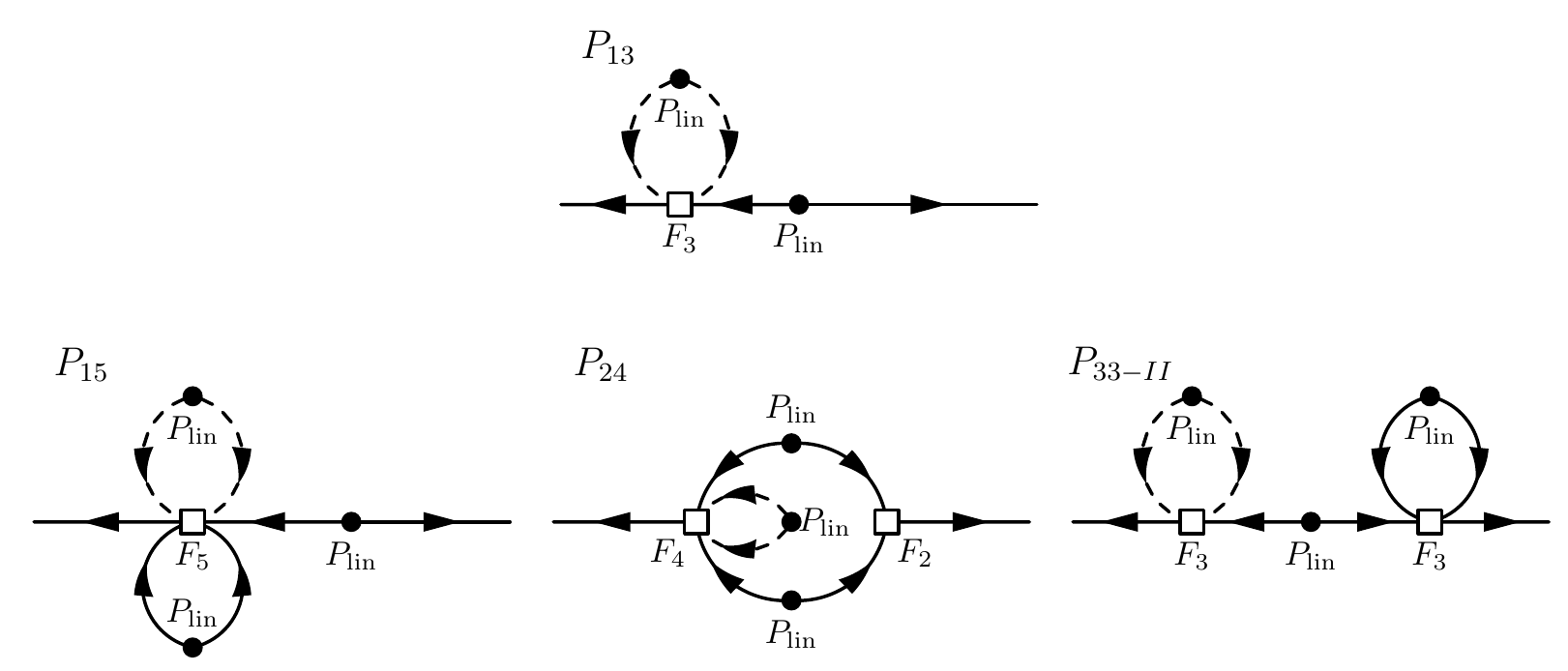}
\caption{Diagrams that are regularized in our approach. The dashed loops are the ones where the momenta is large and are fixed by a counterterm.}
\label{fig:loopdiagreg}
\end{figure}

The first thing one notices is that some of the two-loop diagrams contain inside of them a subdiagram that looks like a one-loop diagram. The EFT procedure  amounts to adding a counterterm that corrects the mistakes introduced when a high momenta is running in a loop. This is schematically shown in Figure  \ref{fig:loopdiagreg}. If at least some of the two-loop diagrams contain pieces that look like one-loop ones, then the same counterterms that have fixed the one-loop subdiagram would fix the two-loop ones. To accomplish this, one would need to solve the equations of motions with the one-loop counterterm as a source to obtain a solution linear in the amplitude of the counterterm but up to cubic in the initial conditions $ \delta_0$. Because the counterterm will be a source in the equations of motion acting over time, carrying out this calculation would require specifying the time dependence of the one-loop counterterm. This program was carried out in \cite{Carrasco:2013mua} as well as in the case of the bispectrum \cite{Angulo:2014tfa,Baldauf:2015qfa}. 

But even for the diagrams that naively look like those in the one-loop calculation, putting the one-loop counterterm into the equations of motion does not necessarily fix all the loops correctly. The point is simple: in the one-loop case, the diagram is computing the effect of a short mode that evolves in a linear long wavelength background. Thus, the time evolution of this background is given by the linear growth factor. In some of the two-loop diagrams the short modes in the loop are evolving in a background that is quadratic or cubic and thus the details of their evolution and the value of the counterterm could not be the same. This fact was already noted in the one-loop bispectrum calculation, where it was shown that the counterterms coming from the time evolution of the linear counterterm are not able to capture the UV-sensitivity of the SPT loops \cite{Baldauf:2015qfa}. 

Of course, in addition to the terms that derive from the linear counterterm through the equations of motion, there are those that arise from the new quadratic and cubic contributions to the stress tensor. Once all of these counterterms are included, one has sufficient freedom to correct all UV mistakes at this order. The entire set of counterterms could be fixed by studying the power spectrum, bispectrum and trispectrum. 

In \cite{Carrasco:2013mua}, the first two-loop calculation in the EFTofLSS, only the counterterms that follow from the leading order one were kept. Thus, the one- and two-loop counterterms depended on only one free parameter (and its time dependence). This was done mainly for simplicity, as one could not fit multiple parameters simultaneously from the available power spectrum data. 
Here, we will take a similar strategy, in that we will also study a one-parameter family of counterterms, but we will fix them in a different way. Because in both, our calculation and in \cite{Carrasco:2013mua}, one is using an ansatz for the two-loop counterterm, one should recognize that in all generality its amplitude could be somewhat different than the one being calculated. 

\subsection{UV-sensitivity at one loop}\label{sec:uv1}
\begin{figure}[t]
\centering
\includegraphics[width=0.49\textwidth]{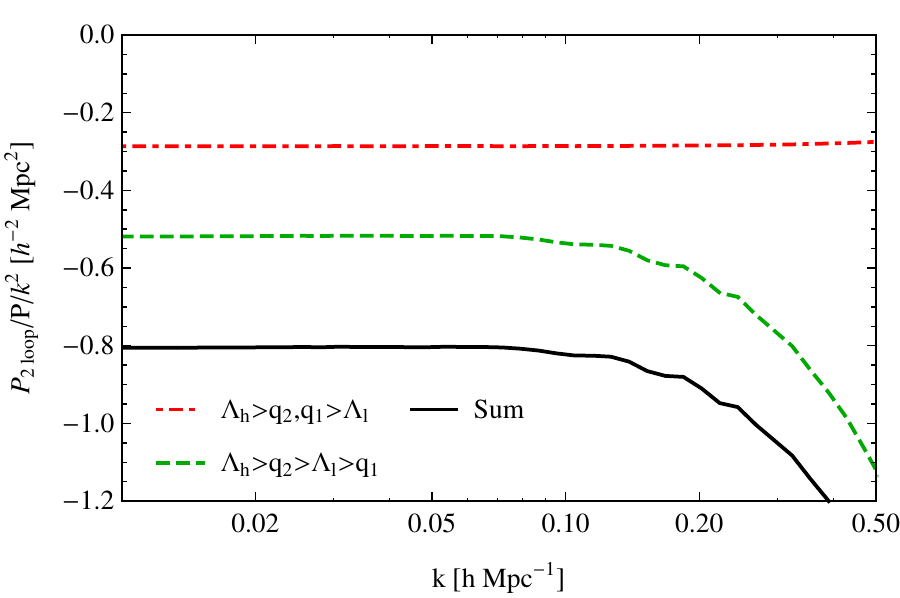}
\includegraphics[width=0.49\textwidth]{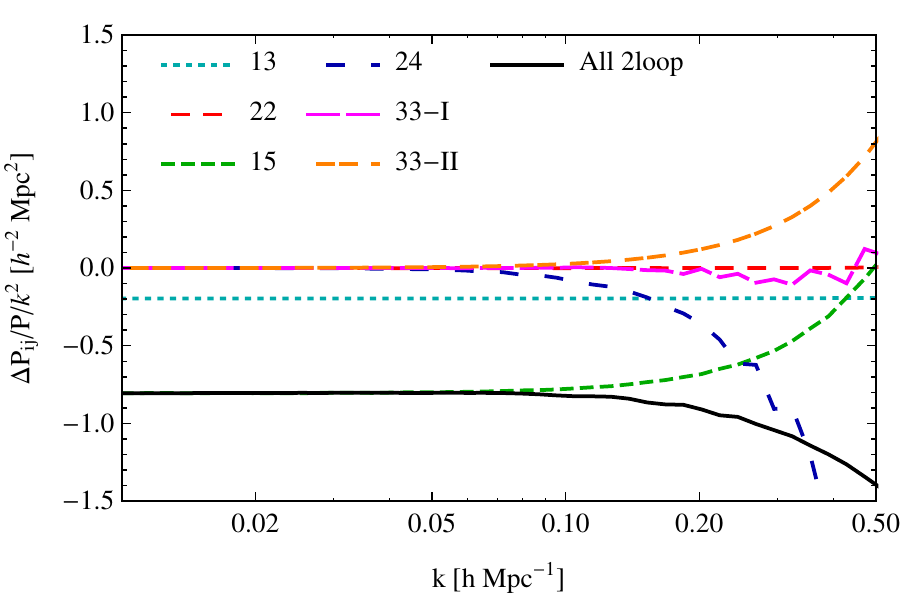}
\caption{Effect of changing the cut-off from $\Lambda_\text{h}=5\ihMpc$ to $\Lambda_\text{l}=1\ihMpc$ for the one and two loop calculations normalized by $k^2 P$. \emph{Left panel: }Contributions from the low-high and high-high terms (single- and double-hard). The mixed term clearly dominates the $k^2 P$ part and also the deviations from this behavior. 
\emph{Right panel: }Contributions from the separate diagrams. At the one loop level $P_{13}$ leads to a $k^2 P$ contributions, whereas the $k^4$ contribution from $P_{22}$ is suppressed.
$P_{15}$ dominates the $k^2 P$ part but for the deviations from this scaling, there is a cancellation between $P_{15}$, $P_{33\text{-II}}$ and $P_{24}$. Like $P_{22}$ in the one loop case, the $k^4$ term arising from $P_{33-I}$ is suppressed.
}
\label{fig:cut-offdep}
\end{figure}

The counterterms introduced in the EFT are there to model the effects of the small scale dynamics on larger scale modes. Thus, a place to look for an ansatz for the relative sizes of the EFT counterterms could be to study the effect of a shell of power at high loop momenta computed in SPT. We can fix the ratio between the various counterterms in the EFT to the one given by this ansatz but leave the amplitude of the small scale power in the shell as one overall free parameter. This ansatz makes the final results insensitive to the small scale power in the SPT calculation.

For this purpose, we start by computing the contribution of a shell in momentum space between $\Lambda_\text{h}=5\ihMpc$ and  $\Lambda_\text{l}=1\ihMpc$ to the one-loop power spectrum in SPT. This choice is somewhat arbitrary, but provides us with a sufficiently significant change to see the effects and furthermore the lower limit is sufficiently far away from the scales of interest $k\approx 0.1\ihMpc$ to warrant a separation of scales.
We call this contribution to the one-loop power spectrum $P_\text{1loop}^\text{sh}=2 P_{13}^\text{sh} + P_{22}^\text{sh}$ and the results are shown in Fig.~\ref{fig:cut-offdep}. We recover the standard result, that for $\Lambda$CDM at small wavenumbers the $P_{13}^\text{sh}$ contribution dominates and scales as $k^2 P_{11}$. The subdominant $P_{22}^\text{sh}$ contribution scales as $k^4$. In this language, what is usually called the $c_s^2$ correction in the EFT is nothing other than the functional form of the $k\rightarrow 0$ limit of $P_{13}^\text{sh}/k^2 P_{11}$.   

The value of $P_\text{1loop}^\text{sh}$ depends not only on the amplitude of the power added on the shell but also  on the position of the shell. In our ansatz for the counterterms, we can also use analytic expressions obtained in the limit that the momentum of the shell ($q_1$)  is much bigger than the momentum of interest $k \ll q_1$. We will call this limit $P_\text{1loop}^{q_1\to \infty}$, basically moving the shell to infinity. For the leading and sub-leading contributions we obtain:
\beq
P_{13}^{q_1\to \infty}= -k^2 P\underbrace{\frac{61}{630} \int _{\vec q}\frac{P(q)}{q^2}}_{l^2\equiv \frac{61}{210}\sigma^2_d}+k^4 P\frac{2}{105}\int _{\vec q} \frac{P(q)}{q^4} + \ldots
\eeq
The leading contribution is proportional to the high-$q$ contribution to the one dimensional displacement dispersion $\sigma_d^2=1/6 \pi^2 \int \derd q\; P(q)$ and the shell power will thus be $P_{13}^\text{sh}=-61/210 k^2 \sigma_{d,\text{sh}}^2 P_{11}$. For the shell under consideration here, we have $\sigma_{d,\text{sh}}^2=0.68 \hMpcsq$.
It is interesting to note that the square of this coefficient has a factor 100 stronger cut-off dependence than the coefficient of the subleading $k^4 P$ contribution, which makes sense since the integral of the latter is suppressed by two additional powers of $q$ in the UV.  Our strategy will be to add the effect of this shell of power computed up to two-loops to the standard SPT results with a free parameter.

This approach is equivalent to the EFT, where the effect of a high-$k$ shell can be captured by the leading counterterm $\cs$ (it should be noted that $\cs$ is not equal to the parameter $d^2$ in Eq.~\eqref{eq:tauLO} as $\cs$ is the result of a time integral over the Green's function and $d^2$). In terms of the standard notation in the literature,
\beq
P_\text{ctr,1loop} \equiv - 2 k^2 c_s^2 P_{11}
\label{eq:oneloopctr}
\eeq
and at the level of the density field it corresponds to the term $\tilde{\delta}_{(1)}=-c_s^2 k^2 \delta_{(1)}$.
It is common practice to fix the coefficient $\cs$ after the full one-loop calculation has been subtracted from the data and we will adopt this convention here. Thus, the true coefficient $\kappa_2$ of the  $k^2 P_{11}$ part of the low-$k$ limit of the data is fixed to be
\beq
\kappa_2=-\frac{61}{105}\sigma_d^2 -2 c_s^2.
\label{eq:kappa2}
\eeq
Consequently, the number $\cs$ effectively contains all the higher order loop contributions to $\kappa_2$, their counterterms and the true small scale contribution. In particular, no higher loop contributions to $\kappa_2$ should be calculated and to the extend that such terms are present in higher order calculations, they should be removed.\\
Numerically, we will find below in accordance with previous studies that $c_s^2$ is a positive number of order $1 \hMpcsq$. This means that we are effectively increasing the power in a high-$k$ shell in perturbation theory, but the effect is an enhanced large scale damping of the non-linear power spectrum.
\subsection{UV-sensitivity at two loops}\label{sec:uv2}
We now  evaluate\footnote{The numerical integrals for the two-loop expressions are performed with the \texttt{CUBA} libraries \cite{Hahn:2004fe} \texttt{SUAVE} routine employing the IR-safe integrand \cite{Blas:2013aba,Carrasco:2013sva}.
} the total two-loop power spectrum as well as its constituent pieces for two different cut-offs $\Lambda_\text{l}=1\ihMpc$ and $\Lambda_\text{h}=5\ihMpc$. 
As we show in Fig.~\ref{fig:cut-offdep}, adding this shell of power primarily affects the $k^2 P_{11}$ coefficient. This piece should be absorbed by the counterterm that was already introduced at one loop in the previous section. The only relevant parts are the deviations from the $k^2 P_{11}$ behavior for $k>0.1\ihMpc$. These are the deviations that we want to capture. These non-trivial pieces should be captured by the two-loop counterterms. 

In contrast to the one-loop calculation, we now have two momenta that are integrated over and thus we have to distinguish two cases: i) both loop momenta are large (both loop momenta in the in the high-$k$ shell, double-hard) or ii) only one loop momentum is large with respect to the other momenta in the problem (one momentum in the high-$k$ shell, single-hard). The left panel of Fig.~\ref{fig:cut-offdep} shows these two contributions separately. We immediately see that the double-hard limit is basically degenerate with the $k^2 P_{11}$ behavior for all the $k$s of interest and thus is not very relevant for our calculation. There is a slight upturn for high wavenumbers that we will discuss in more detail below. The single-hard contribution also has a $k^2 P_{11}$ part, in which we are not interested, but beyond this it has the interesting new scale dependence that should be captured by new counterterms. This motivates us to consider the single-hard limits of the two loop calculation.

The terms leading to the $k^2 P_{11}$ contribution in the shell calculation are also present in the finite part of the two-loop calculation, actually governing its low-$k$ behaviour. If this contribution was kept in the final calculation, it would change the value of the parameter of the one-loop counterterm $\cs$ in Eq.~\eqref{eq:oneloopctr} that was introduced to regularize the $P_{13}$ contribution. As we stated in the previous section, our strategy is to fix this number after the one-loop calculation.
We thus decide to remove the $k^2 P_{11}$ term from the finite part of the two-loop calculation. This can be done by either calculating the limit of $P_{15}$ as we do in Eq.~\eqref{eq:p15doublehard} or by fitting the very low-$k$ limit of the numerical calculation. We will denote the two-loop calculation, from which the degenerate part has been removed as $\bar P_\text{2loop}$. More generally, all terms that have been corrected for degeneracies with lower order counterterms will be decorated with an overbar.\footnote{In \cite{Carrasco:2013mua} the subtraction of the two-loop contribution proportional to $k^2 P_{11}$ was accomplished by introducing the parameter $c_{s,(2)}^{2}$.}

Let us now come back to the limits of the two-loop calculation. Before we discuss the single-hard limits that we deemed responsible for the new counterterms at two loops, we  discuss one double hard limit, 
namely the one of $P_{33-II}$, that leads to
\beq
P_{33-II}^{q_1,q_2\to\infty}=\left(\frac{61}{210}\right)^2 k^4 \sigma_d^4 P
\label{eq:p33IIdoublehard}
\eeq
The counterterm for this term is automatically included once the square of the leading order term at the field level $\langle \tilde{\delta}^{( 1)} \tilde{\delta}^{(1)}\rangle \propto c_s^4 k^4 P_{11}$ is considered. 
There is also a piece of $P_{15}$ that has this same structure. 

As before, we can obtain analytical formulas when the shell is taken to be at infinite momenta. Because we only care about the piece that does not look like $k^2 P_{11}$, we focus on the case when one of the two momenta is running in the loops is in the shell ($q_1$) while the other one remains finite ($q_2$), \emph{i.e.}, the single-hard limit. Fig.~\ref{fig:cut-offdep} shows that the total effect from the two loop terms arises from a cancellation between upturns in $P_{15}$ and $P_{33-II}$ and a downturn in $P_{24}$, while $P_{33-I}$ is basically flat. The overall effect is a residual suppression of power on small scales. We would not have needed to consider an explicit shell, but could have considered all the limits of the two loop calculation (as we do in Appendix~\ref{app:limits}) to see which terms will lead to relevant counterterms. We considered the approach presented here more pedagogical. 

Leaving the double-hard limits and the suppressed single-hard limits for discussion in Appendix~\ref{app:limits}, let us give here only the relevant terms. The formulae for these single-hard limits read:  
\bea
P_{15}^{q_1\to \infty}&=& 2\times5\times 3 \int_{\vec q_2}\frac{3}{127135008000 r_2^5}\Biggl\{-2 r_2 (2266005 - 33470730 r_2^2 + 
     187902172 r_2^4 - 9879110 r_2^6 + 1167375 r_2^8)\nonumber\\ 
&&  + 15 (-1 + r_2^2)^3 (-151067 - 451074 r_2^2 + 77825 r_2^4) \log\left[\frac{
    1+ r_2}{|1 - r_2|}\right]\Biggr\}P(q_2) \sigma_d^2 k^2 P(k)   \;, \label{eq:p15singlehard}\\
P_{24}^{q_1\to \infty}&=&4\times 3\times 
3k^2 \sigma_d^2 \int_{\vec q_2} \frac{(-32879 \mu_2 + r_2^3 (6176 - 48096 \mu_2^2) + 
  32 r_2^2 \mu_2 (1117 + 1503 \mu_2^2) + 
  r_2 (-25933 + 16892 \mu_2^2))}{4074840 r_2 (1 + r_2^2 - 2 r_2 \mu_2)} \nonumber \\
 & &\times F_2(\vec q_2,\vec k-\vec q_2) P(|\vec k-\vec q_2|)P(q_2)  \;,  \label{eq:p24singlehard}\\
P_{33-II}^{q_1\to \infty}&=& -2\sigma_d^2 \frac{61}{210}k^2 P_{13}(k)  \;, \label{eq:p33IIsinglehard}
\eea
where we defined $r_i = q_i/k$ and $ \vec{k}\cdot \vec{q_i} =  \mu_i k q_i$. Note that all three single hard limits are proportional to the high-$q$ contribution to $\sigma_d^2$, \emph{i.e.}, the small scale displacement dispersion.
As for $P_{33-II}^{q_1,q_2\to\infty}$ the counterterm for $P_{33-II}^{q_1\to \infty}$ is automatically included once the leading order counterterm at the field level is correlated with the third order field, leading to $\langle \tilde{\delta}^{( 1)}\delta^{(3)}\rangle \propto -2c_s^2 k^2 P_{13}$.
The $P_{24}^{q_1\to \infty}$ term corresponds to the UV-sensitivity of the bispectrum term $B_{114}$ calculated in \cite{Baldauf:2015qfa} and is thus fixed by the corresponding counterterms. The $P_{15}^{q_1\to \infty}$ term corresponds to a contraction of the UV-limit in the trispectrum term $T_{1115}$ and should thus be regularized by the corresponding counterterm.

Slightly problematically, the $k^2$ part of the $P_{15}^{q_1\to \infty}$ integral is log-sensitive in $q_2$ for $q_2\gg k$
\beq
P_{15}^{q_1\to \infty,q_2\to \infty}=-2\times 5 \times 3\times 3  \sigma_d^2 k^2 P(k)\int_{\vec{q}_2}\frac{120424}{45147375}P(q_2)\; .
\label{eq:p15hardsoft}
\eeq
This limit is proportional to $k^2 P_{11}$ and thus completely degenerate with the leading order counterterm $c_s^2$. It is this limit that leads to the offset in the single-hard limit in the left panel of Fig.~\ref{fig:cut-offdep}. It is continuous with the limit in which in a first step $q_1$ and $q_2$ become large (the double hard limit of Eq.~\ref{eq:p15doublehard}) and then one momentum is taken to be smaller than the other one $q_2\ll q_1$. To the extend that we are not interested in this contribution we define a new limit from which this term is removed at the integrand level
\beq
\bar P_{15}^{q_1\to \infty}= P_{15}^{q_1\to \infty}+2\times5\times 3 \times 3\sigma_d^2 k^2 P(k) \int_{\vec q_2}\frac{120424}{45147375} P(q_2)\; .\label{eq:p15singlehard2}
\eeq
\subsection{Ansatz for the two-loop counterterm}\label{sec:ansatzct}
Our ansatz to have a one parameter counterterm is to use:
\beq
\bar P^\text{sh}= \alpha (P_\text{1loop}^\text{sh} + \bar P_\text{2loop}^\text{sh})\; ,
\eeq
where $\alpha$ is an overall free parameter that can be set by looking at the piece of the power spectrum that scales as $k^2 P_{11}$ at very low $k$ and $\bar P_\text{2loop}^\text{sh}$ is the two-loop power spectrum obtained from the shell calculation with the degenerate $k^2 P_{11}$ piece removed. As we have seen above, all the relevant limits at one- and two-loop are proportional to $\sigma_d^2$, such that we can relate our ansatz to the usual EFT language by setting $\alpha=210 c_\text{s}^2/(61 \sigma_d^2 )$.

More explicitly, for the counterterm at the two loop level we will consider
\bea\label{counter-ansatz1}
P_\text{ctr}&=& \alpha \Bigl[2P_{13}^{q_1\to \infty}+ 2\bar P_{15}^{q_1\to \infty}+2 P_{24}^{q_1\to \infty}+ P_{33-II}^{q_1\to \infty}\Bigr] \nonumber \\
&\equiv& P_\text{ctr,1loop}+P_\text{ctr,2loop}\; .
\eea
All the terms in the right hand side of Eq.~\eqref{counter-ansatz1} are proportional to $\sigma_d^2$, and in fact these are the only terms proportional to $\sigma_d^2$. Hence, our ansatz is nothing other than choosing the value of $\sigma_d^2$ by matching the low $k$ behavior of the power spectrum from simulations to the $k^2 P_{11}$  template. Since this is an important point, let us repeat again the basic idea of our approach. The relation between $\cs$ and $\sigma_d^2$ that is found at the one-loop level in Eq.~\eqref{eq:kappa2} is used in order to cure the UV sensitivity of the two-loop integrals. Effectively, for all occurrences of the problematic $\sigma_d^2$ in the two-loop integrals we add a $\cs$ counterterm. This is what is shown in Eq.~\eqref{counter-ansatz1} and we end up with a one-parameter model for the UV sensitive parts of the one- and two-loop integrals.

Finally, note that the standard IR cancellation when $q_2\ll k \ll q_1$ still happens among the  $q_1$-limits computed above:  $P_{15}$, $P_{24}$ and $P_{33-II}$.  In this case the $1/q_2^2$ motion contributions cancel and only long wavelength tides survive: 
\beq
P_{33-II}^{q_1\to\infty,q_2\to0}=P(k)\int_{\vq_1}\int_{\vq_2}\left[-\frac{3538}{99225 r_1^2} + \frac{61}{1890}\frac{1}{r_1^2 r_2^2}\right]P(q_1)P(q_2) \;,
\eeq
\beq
2P_{24}^{q_1\to\infty,q_2\to0}=P(k)\int_{\vq_1}\int_{\vq_2}\left[-\frac{1361863}{5942475 r_1^2} - \frac{61}{945} \frac{1}{r_1^2 r_2^2}\right]
P(q_1)P(q_2) \;,
\eeq
\beq
2P_{15}^{q_1\to\infty,q_2\to0}=P(k)\int_{\vq_1}\int_{\vq_2}\left[-\frac{902354}{4729725 r_1^2} + \frac{61}{1890} \frac{1}{r_1^2 r_2^2}\right]P(q_1)P(q_2) \;,
\eeq
\beq
2P_{15}^{q_1\to\infty,q_2\to0}+P_{33-II}^{q_1\to\infty,q_2\to0}+2P_{24}^{q_1\to\infty,q_2\to0}=-P(k)\int_{\vq_1}\int_{\vq_2}\frac{12670991}{27810783 r_1^2}P(q_1)P(q_2) \;.
\eeq

\begin{figure}[t]
\centering
\includegraphics[width=0.49\textwidth]{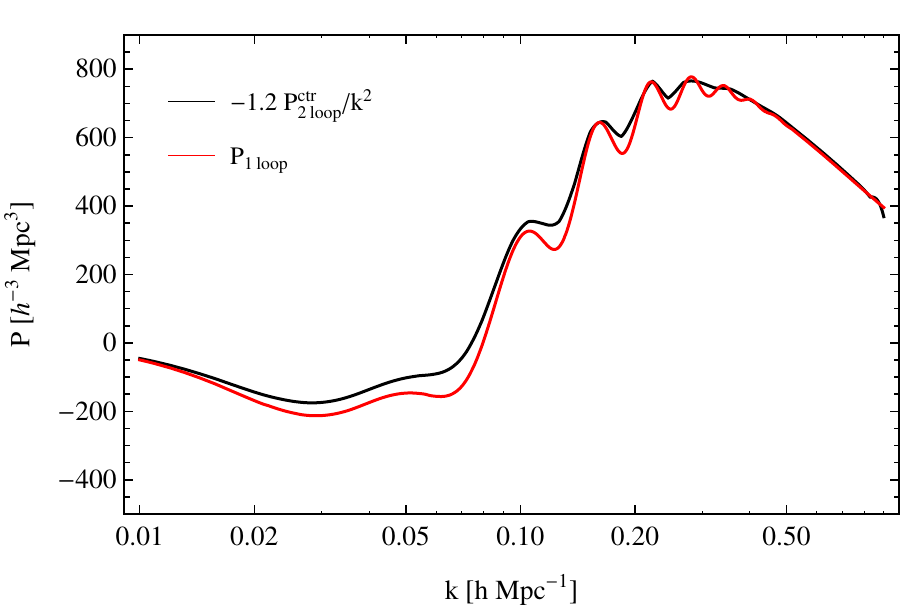}
\caption{Comparison between the two-loop counterterm deduced from the divergencies and the one-loop power spectrum weighted by wavenumber squared. We see that the explicit calculation of the two loop counterterms $\bar P_\text{ctr,2loop}$ is proportional to the naive estimate $k^2 P_\text{1loop}$.}
\label{figansatz}
\end{figure}

\subsection{An even simpler ansatz}\label{sec:simple}
Finally, we could consider what is perhaps the simplest ansatz of all. Just as in the case of the stresses parametrized by $\tau_\theta$ in equation (\ref{nlocounter}), one can parametrize the counterterms in such a way that one of the terms is just proportional to the density computed in SPT. That is the terms relevant for the two-loop calculation at the level of the fields could be written as:
\beq
\delta = \delta_{(1)}+\delta_{(2)} + \delta_{(3)}+\delta_{(4)}+\delta_{(5)} - l^2 \triangle (\delta_{(1)}+\delta_{(2)}+ \delta_{(3)}) + \ldots \;,
\eeq
where the ellipsis account for the terms arising from other quadratic and cubic counterterms. An extremely simple ansatz is to set those additional terms to zero. This would lead to the following expression:
\beq
P_\text{ctr,simple}= - k^2 l^2 \Bigl[P_{11} + P_\text{1loop}\Bigr].
\eeq

As we discussed earlier, neither of these ansatzes is expected to be perfect, and nothing short of fixing all the counterterms by studying the three- and four-point functions or projections at the field level would be perfect. The philosophy of this paper is to write down examples which are expected to have roughly the right size and use those to asses how big these terms are expected to be while keeping in mind the uncertainty in their size. Fig.~\ref{figansatz} compares the two formulae proposed in this section. They are in reasonable agreement in terms of both the expected shape and the size of the correction. One could definitely argue that $P_\text{simple}$ is perhaps too simplified as one is ignoring effects that we know are there and are furthermore comparable to those being included. Our $P_\text{ctr}$ defined in Eq.~\eqref{counter-ansatz1} has all the relevant terms included, although perhaps some of their relative amplitudes are not correct in detail.

\subsection{Relative size of the corrections}

\begin{figure}[t]
\centering
\includegraphics[width=0.49\textwidth]{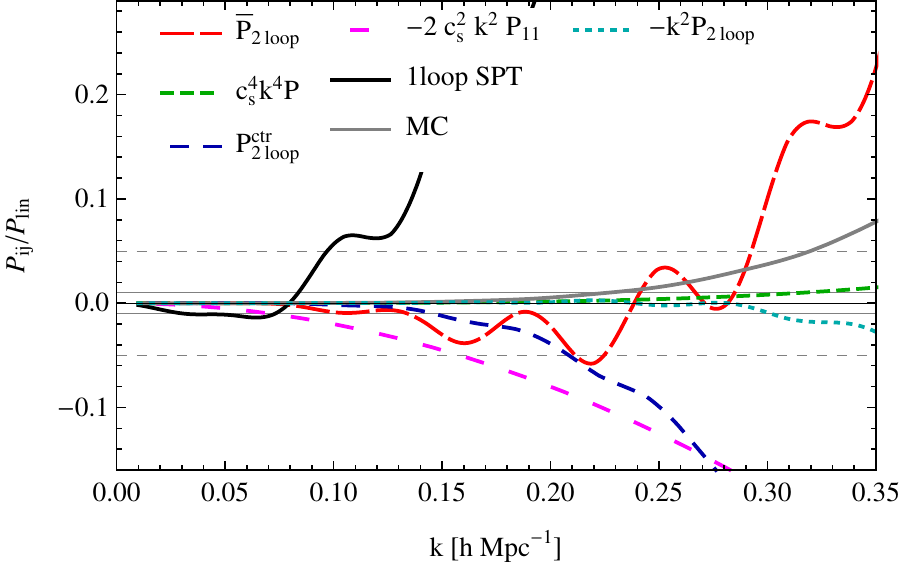}
\includegraphics[width=0.49\textwidth]{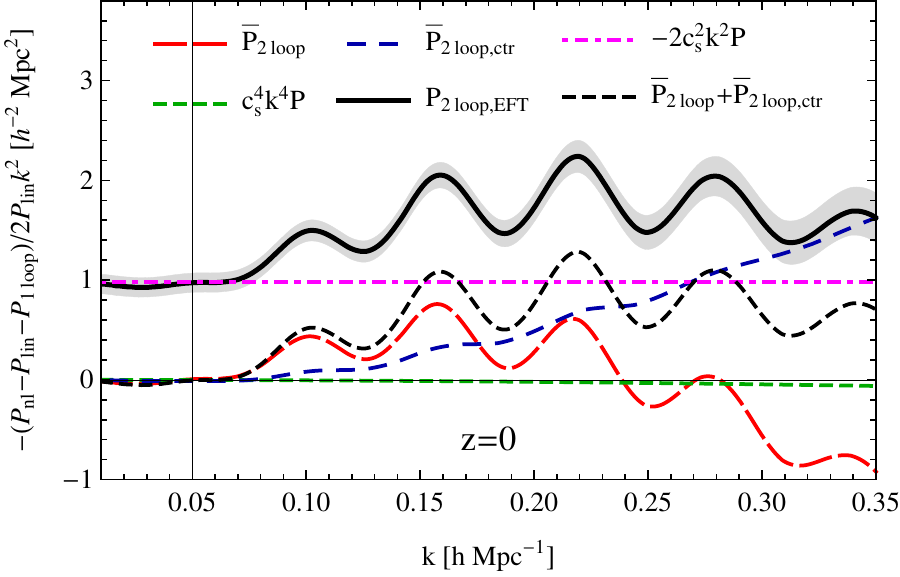}
\vspace{-0.3cm}
\caption{
\emph{Left panel: }Estimate of the size of the corrections arising from various contributions to the two-loop calculation.  The finite part of the two-loop calculation leads to percent level corrections at $k=0.1 \ihMpc$. We also show the corrections from the square of the speed of sound term $k^4 P_{11}$, which is suppressed over the range considered here. The size of the three-loop counterterm can be estimated as $\mathcal{O}(1)\times k^2 \bar P_\text{2loop}$ and leads to percent level contributions at $k=0.3\ihMpc$. We also show the estimate for the stochastic part of the total power spectrum from \cite{Baldauf:2015tlb} which leads to percent level corrections at $k=0.25\ihMpc$.
\emph{Right panel: }Estimator for the leading EFT coefficient $\cs$. The model is evaluated for $\cs=0.98\hMpcsq$ and the gray band shows the effect of a $10\%$ change in this value. Note that at $k=0.2\ihMpc$ the one-loop counterterm and the two-loop correction are of the same order. The two-loop term leads to a considerable scale dependence of $\hat c_s^2$ for $k>0.07\ihMpc$.}
\label{fig:csscaldep}
\end{figure}

Before comparing with simulations, we can take our model for the power spectrum and calculate the sizes of the different terms. This comparison will allow us to estimate how well we expect our formulae to agree with simulations  and estimate reach of perturbation theory. In particular, now that we have an estimate of the two-loop terms and their associated counterterms we can ask when they make a difference relative to the one-loop terms and ask over what range of $k$s it would be safe to fit for $\cs$ when doing a one-loop calculation only. Our full two-loop EFT power spectrum is 
\beq
P = P_{11} + P_\text{1loop} + P_\text{ctr,1loop}+ \bar P_\text{2loop}+P_\text{ctr,2loop}.
\eeq
We can now compute two quantities:
\bea
\frac{P}{P_{11}} -1&=& { P_\text{1loop}+ P_\text{ctr,1loop}+ \bar P_\text{2loop} +P_\text{ctr,2loop}  \over P_{11}} \nonumber \\
-\frac{P-P_{11} -P_\text{1loop}}{2k^2 P_{11}} &=& c_s^2 - \frac{\bar P_\text{2loop} + P_\text{ctr,2loop} } {2k^2 P_{11}}. 
\eea
The first of these two quantities indicates the size of the various terms as contributions to the power spectrum, the second indicates the relative correction they would make to a fit of $c_s^2$ after subtracting the explicit one loop SPT calculation from the data. We show these quantities in Figure \ref{fig:csscaldep}. The left panel shows that both  $\bar P_\text{2loop}$ and $\bar P_\text{ctr,2loop}$  make roughly a 5\% correction to the power around $k=0.2\ihMpc$. Given that   $P_\text{ctr,2loop}$ is uncertain because we have not used three- and four point function measurements to obtain its amplitude but only have an ansatz, it is difficult to imagine that one could be more accurate than about one percent on these scales. The counterterm is relatively steep, so even though it contributes 5\% around $k=0.2\ihMpc$ at $k=0.5 \ihMpc$ is makes an order unity contribution. In the same panel, we also show the effect of the $c_s^4 k^4 P_{11}$ correction, which is at the sub percent level for the wavenumbers considered here. Note however, that the coefficient of this term should be fitted independently, since it has to capture the subleading UV sensitivities in $P_{13}$ and $P_{15}$ that we have neglected so far. We also estimate the three loop counterterm at the basis of our most simple counterterm ansatz, \emph{i.e.}, we consider it to be given by $-k^2 \bar P_\text{2loop}$. This term leads to percent level corrections at $k=0.3 \ihMpc$, so we should be worried about similarly large corrections from the three loop calculation for even larger scales. Finally, we overplot the size of the stochastic term estimated in \cite{Baldauf:2015tlb}. Given that it leads to percent corrections at $k=0.25 \ihMpc$, we should not expect any perturbative approach to match the full power spectrum to a better accuracy than this. Actually, the perturbative/deterministic calculation performed here should describe the non-linear power from which the stochastic part has been removed.

As we show in the right panel of Fig.~\ref{fig:csscaldep}, when fitting for $c_s^2$, the combination  $\bar P_\text{2loop} + P_\text{ctr,2loop}$ changes  $c_s^2$ by about 50\% between $k=0.05\ihMpc$ and $k=0.20\ihMpc$. About half of this change is from the finite part of the two-loop calculation and half from the counterterms. These two corrections are of the same amplitude at $k=0.18\ihMpc$. Due to the presence of these corrections, a measurement of $\cs$ without consideration of the two-loop terms is not possible for wavenumbers exceeding $k=0.07 \ihMpc$. Besides the broadband upturn, there are also considerable wiggles from the Baryon Acoustic Oscillations (BAO) in the finite part of the two loop calculation.

At this point it is perhaps instructive to write an equation relating the change in the inferred value of $c_s^2$ ($\Delta c_s^2$) to changes or errors in the power spectrum $(\Delta P)$:
\beq
\Delta c_s^2 = {\Delta P \over P} {1 \over 2 k^2} \sim {{\Delta P/ P} \over 2 \%} \left({k \over 0.1 \ihMpc}\right)^{-2} \hMpcsq \sim 
{{\Delta P /P} \over 0.2 \%} \left({k \over 0.03 \ihMpc}\right)^{-2} \hMpcsq\; .
\eeq
For values of $c_s^2$ around $1 \hMpcsq$ and a measurement at $k \sim 0.1 \ihMpc$, an accurate measurement of $c_s^2$ requires one to model all other contributions to the power spectrum at the sub-percent level. Besides that, the statistical error should also be at this level. State of the art simulation codes and reasonable simulation volumes can deliver this level of accuracy and precision. 
However, our estimates above show that at this scale, one needs to include the two-loop terms. If one goes to $k \sim 0.03 \ihMpc$ higher loop contributions are negligible, but as we discuss later the required $10^{-3}$ level numerical precision might be challenging.

%
\begin{figure}
\centering
\includegraphics[width=0.49\textwidth]{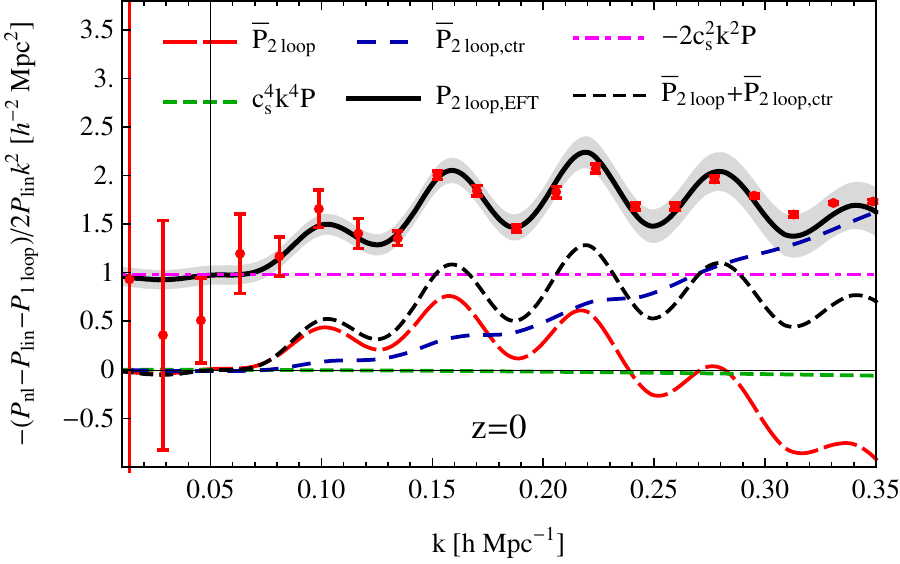}
\includegraphics[width=0.49\textwidth]{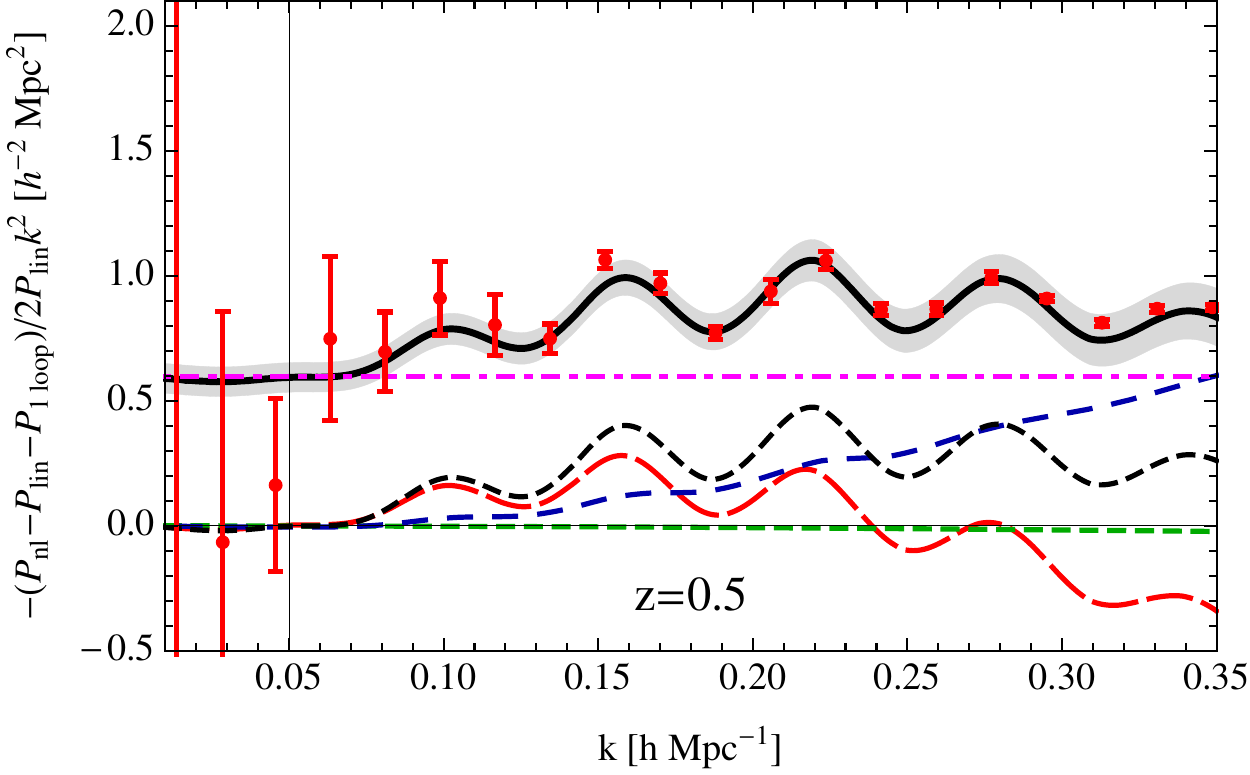}
\includegraphics[width=0.49\textwidth]{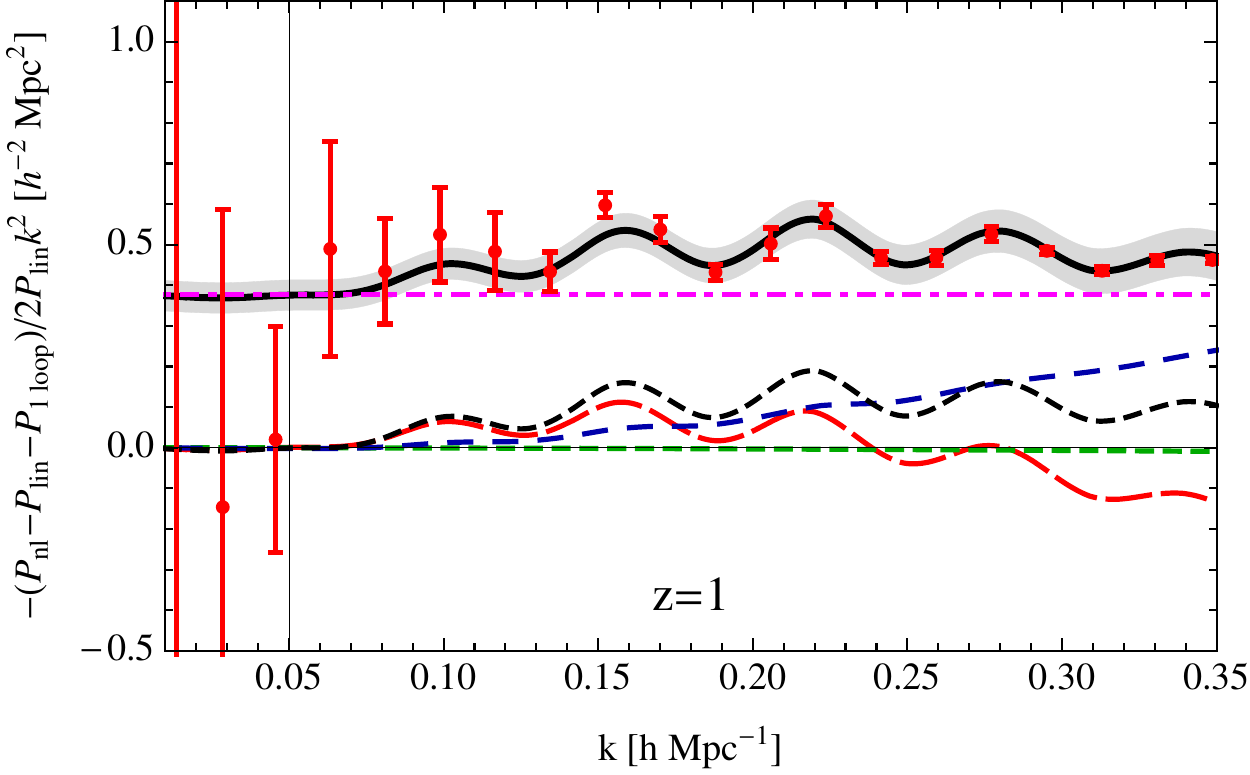}
\includegraphics[width=0.49\textwidth]{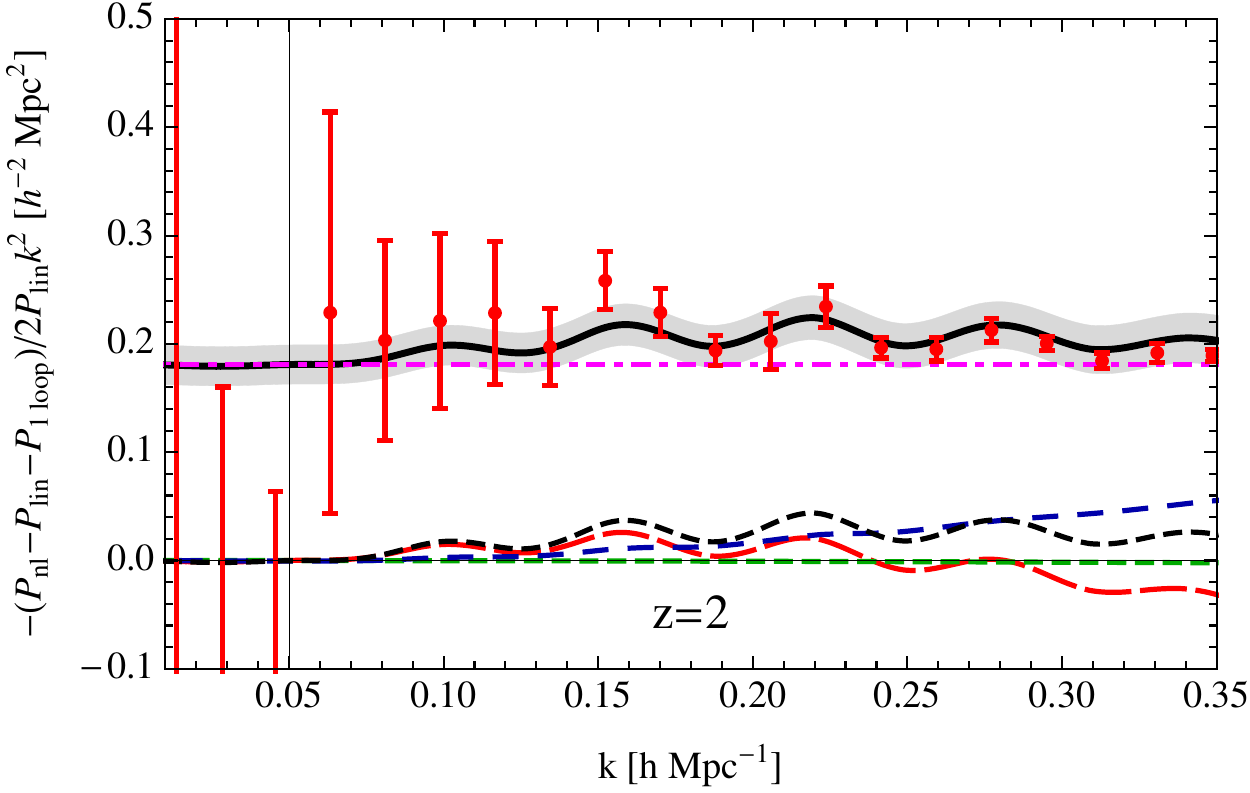}
\caption{
One loop $\cs$ estimator at $z=0,0.5,1,2$ from top left to bottom right before IR resummation. At all redshifts, we see clear evidence for a running of $\cs$ that is described by the scale dependence of the two loop correction and the two loop counterterm.
}
\label{fig:fitcs}
\end{figure}

\section{Comparison with simulations} \label{sec:sims}

As a benchmark for the performance of the perturbation theory we employ a suite of dark matter only simulations of the WMAP7 cosmology \cite{WMAP7} ($\Omega_\text{m}=0.272$, $\Omega_\Lambda=0.728$, $n_\text{s}=0.967$, $\sigma_8=0.81$).
We have run 16 simulations with a box length of $1500 \hMpc$ (L simulation) and also one realization of a smaller size, higher resolution box with $500\hMpc$ box length (M simulation). The simulations are initialized with the second order Lagrangian Perturbation Theory code \texttt{2LPT} \cite{Crocce:2006ve} at redshift $z_\text{i}=99$ and the $1024^3$ particles are subsequently evolved using \texttt{GADGET2} \cite{Springel:2005mi} to redshift $z=0$. For more details on the simulations and some convergence tests see \cite{Baldauf:2015tla}.

For the speed of sound in the one-loop EFT calculation we employ the following estimator
\beq
\hat c_s^2= - \frac{P_\text{nl}-P_{11} -P_\text{1loop}}{2k^2 P_{11}}\; ,
\label{eq:csestimatordd}
\eeq
where $P_\text{nl}$ is the power spectrum from the simulations. 
In Fig.~\ref{fig:fitcs} we show the measurements at redshifts $z=0,0.5,1,2$ from our simulations. The data clearly show a scale dependence with significant deviations from the low-$k$ limit at higher wavenumbers. There are also distinct BAO wiggles in the measurement that have been noted in the literature \cite{Manzotti:2014loa}.
We have corrected the data for $2 \times 10^{-4}$ level deviations in the linear growth factor, that are likely connected to the integration accuracy in \texttt{GADGET} and would lead to a low-$k$ upturn in this figure. Furthermore, we have cancelled the leading order cosmic variance, by actually considering the ratio of non-linear power spectrum and linear (initial) power spectrum measured in the simulations.
The one-loop EFT model (horizontal magenta dashed line) fails to describe the data for $k>0.07\ihMpc$, but the two-loop corrections can explain the residual scale dependence. We find $\cs \approx 0.98 \hMpcsq$ and interpret the difference from previous measurements $\cs \approx 1.6\hMpcsq$ \cite{Carrasco:2013sva}  extracted from the $k=0.15-0.25 \ihMpc$ range as resulting from the two loop contributions. It is also worth noting that the two loop calculation is already doing a very good job at tracking the BAO oscillations, at least for $k< 0.2 \ihMpc$. The calculation based on the UV-limits assumes a time dependence of $\cs$ that matches the one of the SPT term that it is regularizing, \emph{i.e.}, $D^2(a)$. We are using this time dependence to scale our $z=0$ fit to higher redshifts and find very good performance both for the small wavenumber behaviour as well as the scale dependence at higher wavenumbers. All of the redshifts show slightly low datapoints at $k=0.03 \ihMpc$ and $k=0.045 \ihMpc$, that spoil a nice asymptotic behaviour at low wavenumbers that one would expect in the EFT. As we describe in App.~\ref{sec:propa}, this systematic effect goes away if the theory is calculated on the simulation grid, effectively using the same modes that are present in the simulations. 

The discussion of the relative difference between simulation and analytic calculation for the power spectrum itself will be deferred until we discuss the IR-resummation below in Sec.~\ref{sec:irresum}, but the anxious reader might want to look at Fig.~\ref{fig:irresum}. The two-loop calculation agrees with the data at the sub-percent level all the way to $k= 0.3 \ihMpc$ at $z=0$.

\subsection{Time Derivative and momentum correlators at two loops}
\begin{figure}
\includegraphics[width=1.0\textwidth]{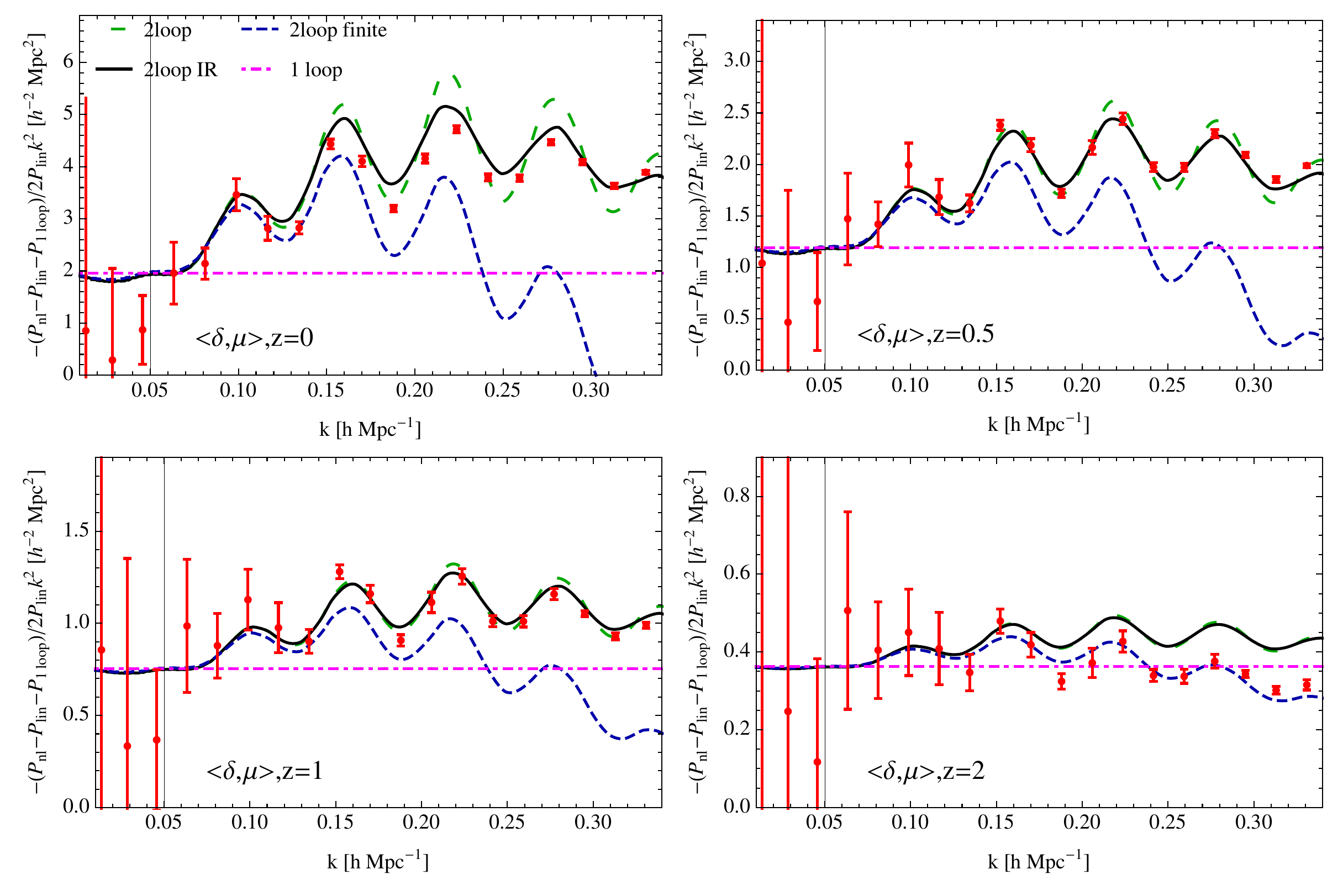}
\vspace{-0.8cm}
\caption{Constraints on the speed of sound and its time dependence from the momentum-density cross power spectrum.}
\label{fig:csdmu}
\end{figure}
\begin{figure}
\includegraphics[width=1.0\textwidth]{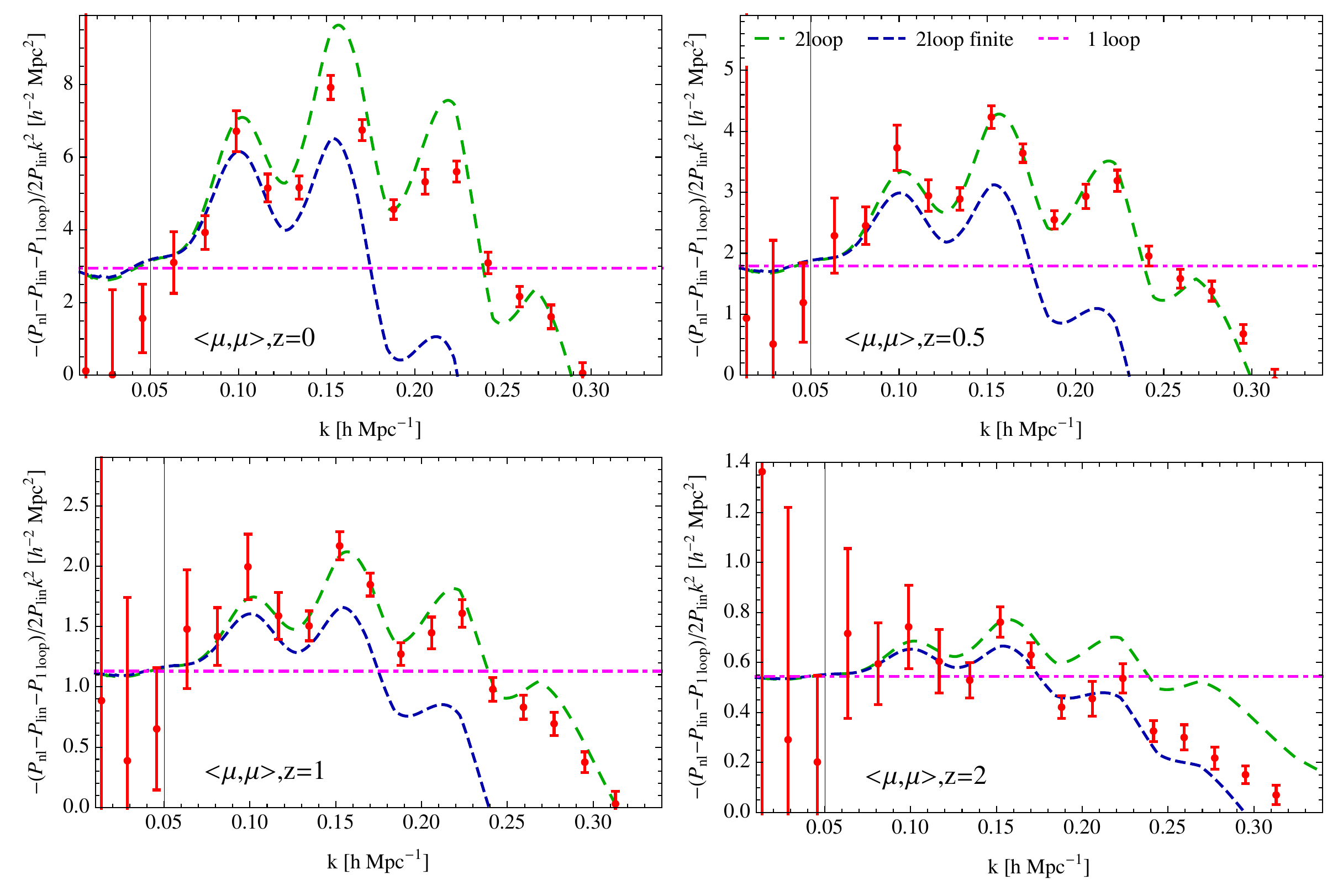}
\caption{Constraints on the speed of sound and its time dependence from the momentum-momentum auto power spectrum.}
\label{fig:csmumu}
\end{figure}

So far we have concentrated at a single observable, the density power spectrum, we would now like to extend the calculation to momentum statistics. This extension is motivated by the fact that the momentum statistics are sensitive to the time dependence of the speed of sound and more sensitive to loop corrections than the density power spectrum itself. We consider momentum $\mu=-\vec \nabla \cdot \left[ (1+\delta)\vec v\right]$ instead of velocity, since the latter is only defined at the particle locations in the simulation and thus the only quantity that can be reliably measured.
To compute correlations involving momenta we then use the continuity equation:
\beq
\mu(\vec k,a)=\delta'(\vec k,a)\; .
\eeq
For the SPT predictions of the momentum-density cross power and momentum-momentum auto power we have
\begin{align}
P_{\delta\mu}=&f\mathcal{H} \Bigl[P_{11}+2 \left(2 P_{13}+P_{22}\right)+3 \left(2 P_{15}+2 P_{24}+P_{33}\right)\Bigr]\; ,\label{eq:delmu}\\
P_{\mu\mu}=&f^2\mathcal{H}^2\Bigl[P_{11}+2 \left(3 P_{13}+2 P_{22}\right)+10 P_{15}+16 P_{24}+9 P_{33}\Bigr]\; .
\label{eq:mumu}
\end{align}
Note that while $P_{\delta \mu}$ is IR-safe, i.e., the cancellation of IR modes between $P_{13}$ and $P_{22}$ still happens (and similarly for the two loop contribution), this is not the case for $P_{\mu \mu}$. 
The density-momentum cross correlation is nothing than the time derivative of the density power spectrum
\beq
\la \delta(\vec k) ,\mu(\vec k')\ra=\la \delta(\vec k) ,\delta'(\vec k')\ra=(2\pi)^3 \ddir(\vec k+\vec k')\frac{1}{2}P_{\delta,\delta}'(k)\; .
\eeq
Thus, if we manage to capture the time dependence of the perturbative corrections to the density power spectrum, we should also be able to describe the density-momentum cross power spectrum at intermediate times.
We immediately see that the importance of loop corrections is enhanced with respect to the density-density power spectrum and thus expect a steeper running of the speed of sound corrections.

Let us now consider the counterterm for the momentum statistics. We will parametrize the time dependence of the counterterm as a power law in linear growth $D$ as $\cs=c_{s,0}^2 D^\gamma$ and with $\delta(\vec k,a)=\delta^{(1)}(\vec k,a)+c_{s,0}^2 D^\gamma(a) \delta^{(1)}(\vec k,a)$ we have then
\beq
\mu(\vec k,a)=\delta'(\vec k,a)=f\mathcal{H}\left[\delta^{(1)}(\vec k,a)+(1+\gamma)c_{s,0}^2 D^\gamma(a) \delta^{(1)}(\vec k,a)\right]
\eeq
In analogy to Eq.~\eqref{eq:csestimatordd} we consider the estimator for the sound speed that first removes the one-loop corrections and maps the residual on the leading order counterterm
\beq
-\frac{P_{(\delta\delta,\delta\mu,\mu\mu)}-P_{11}-P_\text{1loop}}{2 k^2 P_{11}}=\left(c_{s,0}^2, c_{s,0}^2 \frac12(2+\gamma),c_{s,0}^2(1+\gamma)\right)
\eeq
The leading UV-sensitivity of SPT suggests that $\cs$ scales as $\sigma_d^2$, i.e., as $D^2$ ($\gamma=2$). We saw above that this time dependence provided a decent description of the density power spectra.
The constraints from the momentum-density and momentum-momentum statistic are shown in Figs.~\ref{fig:csdmu} and \ref{fig:csmumu}, respectively.
As above, we again see that $\gamma=2$ performs very well both on large scales, but the errors are considerable and $\cs$ constraints for the momentum statistics show a strong scale dependence, as they did for the density.

We will thus continue to consider the the $\gamma=2$ scale dependence and calculate the two-loop momentum counterterms. The counterterm for the density-momentum correlator can be obtained from the time derivative of the density-density counterterm or equivalently from taking the respective limits of Eq.~\eqref{eq:delmu}
\bea
P_{\delta\mu,\text{ctr}}&=& \alpha \Bigl[2P_{13}^{q_1\to \infty}+ 3(2\bar P_{15}^{q_1\to \infty}+2P_{24}^{q_1\to \infty}+ P_{33-II}^{q_1\to \infty})\Bigr] \nonumber \\
&\equiv& P_{\delta\mu,\text{ctr,1loop}}+P_{\delta\mu,\text{ctr,2loop}}=2P_{\text{ctr,1loop}}+3P_{\text{ctr,2loop}},
\eea
We see in Fig.~\ref{fig:csdmu} that this counterterm in combination with the finite part of the two loop calculation can indeed describe the scale dependence of the $\cs$ estimator for $k>0.07\ihMpc$. The SPT inspired time dependence seems to work for this scale dependent part up to $z=1$, but at $z=2$ there are some deviations at $k>0.2\ihMpc$. 
We also overplot the effect of IR-resummation on the two-loop result, which will be described in more detail in the next section.

To obtain the expression for the momentum-momentum correlator one would need to write the expressions for the two-loop counterterms at the field level and take the time derivatives. Here we rather follow the simple approach and consider the limits of Eq.~\eqref{eq:mumu} to obtain
\bea
P_{\mu\mu,\text{ctr}}&=& \alpha \Bigl[6P_{13}^{q_1\to \infty}+ 10\bar P_{15}^{q_1\to \infty}+16 P_{24}^{q_1\to \infty}+9 P_{33-II}^{q_1\to \infty}\Bigr] \nonumber \\
&\equiv& P_{\mu\mu,\text{ctr,1loop}}+P_{\mu\mu,\text{ctr,2loop}}\; .
\eea
The $\cs$ constraints from this statistic are shown in Fig.~\ref{fig:csmumu}. While the agreement between the scale dependence of the two-loop calculation and the data is somewhat worse than for the density and density-momentum cross correlation, the model is still able to roughly capture the broadband scale dependence and the wiggles in the $\cs$ estimator.

%
\begin{figure}[t]
\centering
\includegraphics[width=0.49\textwidth]{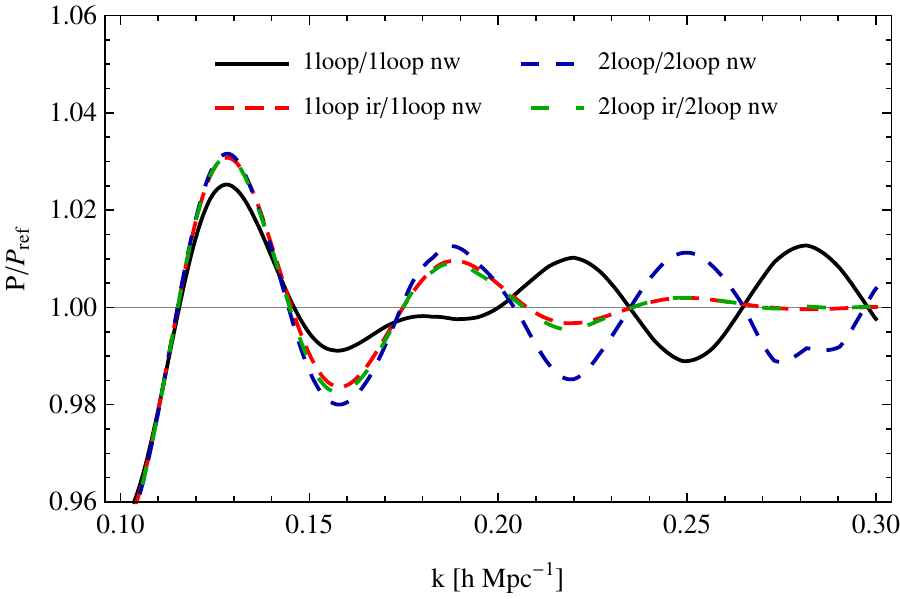}
\caption{Effect of the IR-resummation on the one- and two-loop power spectra. We show the ratio of the power spectra before and after IR-resummation with respect to the corresponding no-wiggle power spectrum in order to remove broadband effects. Below $k=0.2\ihMpc$ the bare two-loop calculation agrees with the IR-resummed one loop calculation at the percent level. After IR-resummation the wiggle part of the one- and two-loop calculation agree, which tells us that the IR-resummation captured the relevant terms in the explicit two loop calculation correctly.}
\label{fig:irresumappr}
\end{figure}

\subsection{IR resumation}\label{sec:irresum}
The EFT corrections discussed so far mainly address the broadband, \emph{i.e.}, short scale behaviour of the power spectra. Another set of corrections that are not fully captured by SPT are the long wavelength motions, which however do not affect the broadband behaviour due to the equivalence principle. The most prominent effect of the long modes is to damp the BAO oscillations. Lagrangian Perturbation Theory captures the effects of these motions better, since it keeps them resummed in an exponential and thus captures their effects to higher orders than the one to which the displacement field has been calculated. SPT only keeps motions to the order explicitly considered in the calculation. An effective way to combine the merits of both approaches is the so called infrared-resummation (IR-resummation) \cite{Senatore:2014via}, which calculates the broadband in SPT and corrects the result to account for the IR-motions.\\
In this section we implement the simple IR-resummation described in \cite{Baldauf:2015xfa}. 
This method multiplies the oscillatory part of the power spectrum (the wiggle part $P_\text{w}$) by an exponential damping but leaves the broadband part (the no-wiggle part $P_\text{nw}$) unaffected
\beq
P_\text{IR}=e^{-\Sigma_\epsilon^2(k)k^2}\Bigl[\left(1+\Sigma_\epsilon^2(k)k^2\right)P_{11,\text{w}}+P_\text{1loop,w}\Bigr]+P_{11,\text{nw}}+P_\text{1loop,nw}\; ,
\eeq
where 
\beq
\Sigma^2(k)=\frac{1}{3}\int_0^k \frac{\derd^3 q}{(2\pi)^3} \frac{P(q)}{q^2}\bigl[1-j_1(q r_\text{BAO})+2j_2(q r_\text{BAO})\bigr] \;,
\eeq
with $j_n$ being the $n^\text{th}$ order spherical Bessel function. We only consider the smoothing due to motions arising from scales much larger than the scale under consideration, for definiteness we choose $\Sigma_\epsilon(k)=\Sigma(k/2)$.

The explicit-two loop SPT calculation contains the resummed motions up to second order, such that the resummation needs to start from third order, \emph{i.e.}, $\Sigma^6$.
\beq
\begin{split}
P_\text{IR}=&e^{-\Sigma_\epsilon^2(k)k^2}\Biggl[\left(1+\Sigma_\epsilon^2(k)k^2+\frac{1}{2}\Sigma_\epsilon^4(k)k^4\right)P_\text{11,w}+\left(1+\Sigma_\epsilon^2(k)k^2\right)P_\text{1loop,w}+\bar P_\text{2loop,w}\Biggr]\\
&+P_{11,\text{nw}}+P_\text{1loop,nw}+\bar P_\text{2loop,nw}\; .
\end{split}
\eeq

The effects of IR resummation are highlighted in Fig.~\ref{fig:irresumappr}. Performing the IR resummation on the bare one-loop calculation leads to considerable changes to the power spectrum. Below $k\approx 0.2 \ihMpc$, the not IR-resummed two-loop calculation performs almost as well as the IR-resummed one loop calculation. The IR-resummation of the two-loop calculation only matters at the percent level for $k>0.2 \ihMpc$.

\begin{figure}[t]
\centering
\includegraphics[width=0.49\textwidth]{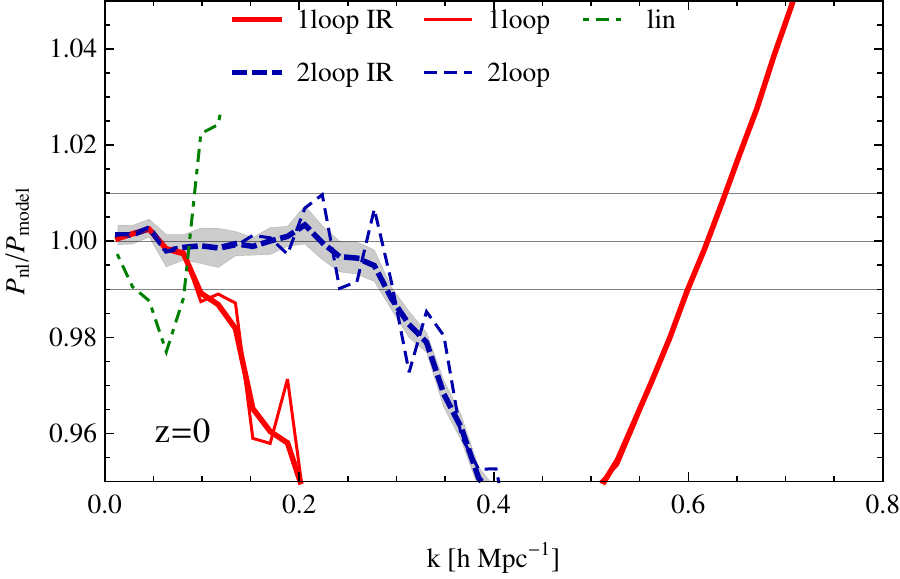}
\includegraphics[width=0.49\textwidth]{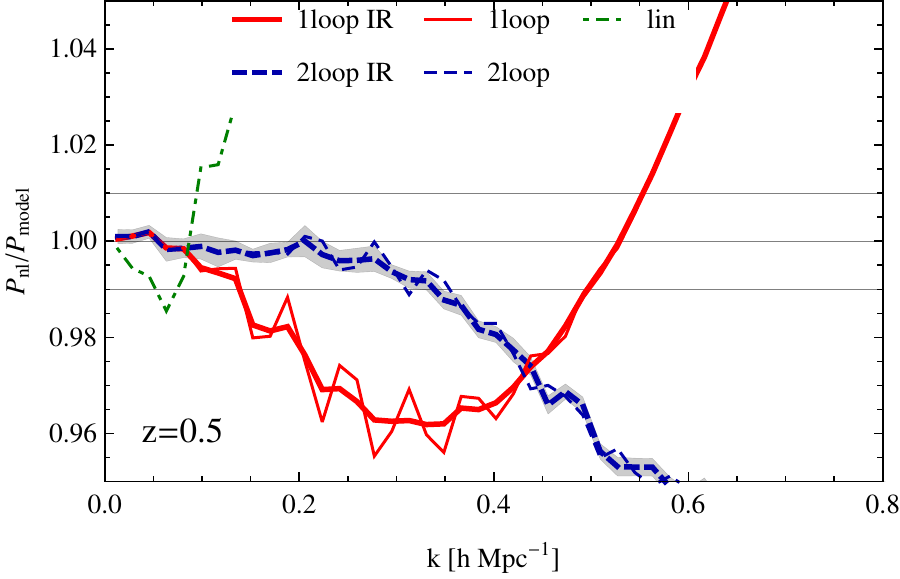}
\includegraphics[width=0.49\textwidth]{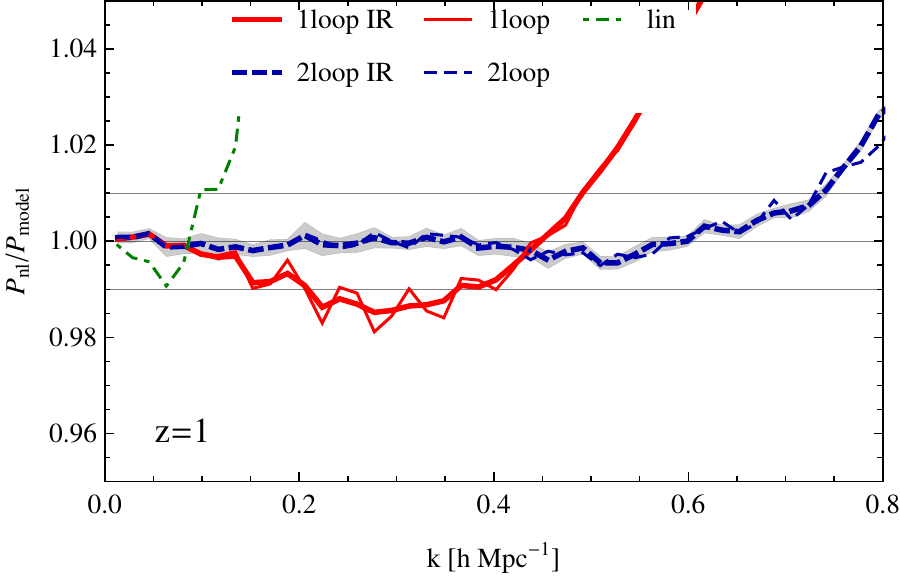}
\includegraphics[width=0.49\textwidth]{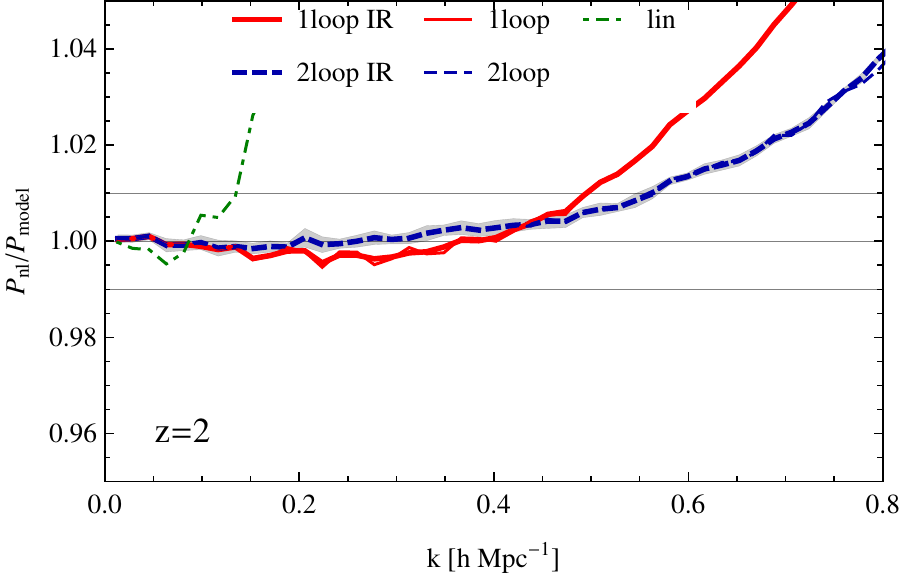}
\vspace{-0.2cm}
\caption{Ratio of the data to the various PT models at redshifts $z=0,0.5,1,2$ from top left to bottom right. We show the linear theory calculation (green dot-dashed), the one-loop EFT (red solid) and the two-loop calculation (blue dashed). For the EFT calculation we show results both before (thin) and after IR-resummation (thick). The ratio is evaluated at the simulation data points and the two sigma errors on these data points are indicated by the gray band.}
\label{fig:irresum}
\end{figure}

As we have seen above in Fig.~\ref{fig:fitcs}, the two-loop calculation is tracking part of the BAO wiggles in the power spectrum residuals after the one-loop result has been removed. Let us now study its performance at higher wavenumbers and in the power spectrum itself. In Fig.~\ref{fig:irresum} we show the performance of the IR-resummed and not IR-resummed one- and two-loop EFT calculations with respect to the non-linear power spectrum extracted from the $N$-body simulation. Let us first discuss the broadband performance. At redshift $z=0$ the one loop calculation extends the range of validity\footnote{For the sake of definiteness we will commonly consider $1\%$ deviations from the theory as the threshold for the range of validity. Many applications will require tighter errorbars on large scales to fix the amplitude. On smaller scales we will anyways suffer from baryonic effects and significant covariance, such that less restrictive requirements could be employed.} of linear theory from $k\approx 0.05 \ihMpc$ to $k\approx 0.1 \ihMpc$. This is significantly less, than usually considered for the range of validity of the EFT at redshift $z=0$ and arises from the fact that we have fixed the leading order counterterm in a way that is compatible with the largest available scales. We then use this parameter to calculate the two loop counterterm. This term, together with the finite part of the regularized two-loop calculation allows us to extend the $1\%$ agreement range to $k\approx 0.3 \ihMpc$. Here we should stop for a minute and reconsider the goal of this exercise. Usually one tries to fit the non-linear power spectrum as well as possible up to the highest possible wavenumber. But actually this should not be the goal of the fit with the deterministic part of the EFT, which we are computing here. The non-linear power spectrum is the sum of this deterministic part and the stochastic part. As we have pointed out in \cite{Baldauf:2015tla,Baldauf:2015tlb}, this stochastic term amounts to a percent of the total power at $k=0.25 \ihMpc$ (and about $3\%$ at $k=0.3\ihMpc$). This means that the deterministic part of the power spectrum should deviate from the non-linear power spectrum by at least this much for $k>0.25 \ihMpc$. The deterministic EFT calculation (performed here) should asymptote to the perturbative/deterministic part of the power spectrum $P_\text{PT}=P_\text{nl}-P_\text{stoch}$ and not to the non-linear power spectrum itself. 
Thus, we have slightly underfitted the $\cs$ parameter, overfitted the power.
Once our EFT calculation is failing, it predicts more power than the non-linear power spectrum (downturn in Fig.~\ref{fig:irresum}). This failure would have happened even earlier and more steeply if the (positive) stochastic contribution had been subtracted from the non-linear power spectrum. A slight increase of $\cs$ would bring our curve closer to the deterministic part of the power spectrum. Furthermore, looking at Fig.~\ref{fig:csscaldep} , we see that both the one- and two-loop terms and their counterterms act to decrease power for a while before crossing zero and adding power. We might thus expect the higher loop calculation to reduce power in the same way at $k > 0.25 \ihMpc$ such that the prediction matches the deterministic part of the power spectrum for these wavenumbers.
\\
Having said this, let us now consider the effect of IR-resummation. As we have seen before, the two-loop results before and after IR resummation agree for $k<0.2\ihMpc$. For higher wavenumbers the IR-resummation indeed reduces the amplitude of the residual BAO wiggles even up to scales where the EFT broadband significantly deviates from the non-linear power.

\section{Outlook to higher orders}
While the range over which the perturbative or deterministic part of the EFT can describe the full non-linear structures is certainly limited by stochastic terms, there might be some hope for further leverage in either precision at low wavenumbers or reach at intermediate wavenumbers from going to higher orders for the deterministic part. As shown in \cite{Blas:2013aba}, a three-loop calculation is in principle feasible, but not very useful when taken at face value. As we have seen above, part of the higher loop calculation is degenerate with lower order counterterms. At the three-loop level this amounts to identify the terms that have already been taken care of by the terms that lead to either $P_\text{1loop,ctr}$ or $P_\text{2loop,ctr}$. 

\begin{figure}
\includegraphics[width=0.7\textwidth]{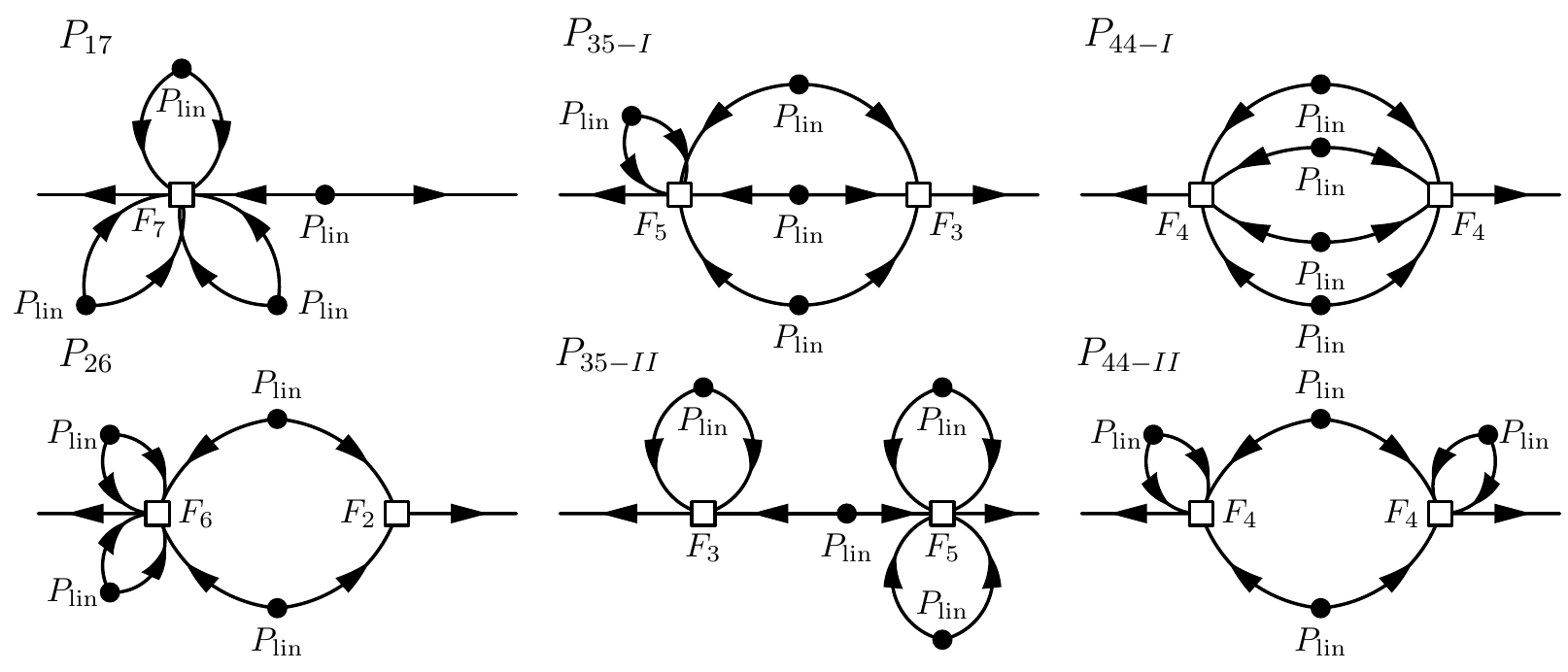}
\caption{Diagrams for the three loop calculation.}
\label{fig:threeloopdiag}
\end{figure}

The three-loop power spectrum is given by
\beq
P_\text{3loop}=2P_{17}+2P_{26}+2P_{35-I}+2P_{35-II}+P_{44-I}+P_{44-II}\; ,
\eeq
with constituent diagrams given in Fig.~\ref{fig:threeloopdiag}.

Let us discuss these terms separately:
\begin{itemize} 
\item $P_{17}$: This is a propagator term. Its leading order contribution for low external wavenumbers scales as $k^2 P_{11}$ and is thus degenerate with the leading order counterterm. To the extent that $\cs$ has been fixed, we can safely subtract this term from the three-loop calculation. We expect such a contribution from the triple hard limit. Then there will be double hard limits, where one of the loops remains at the same order as the external momentum. These terms have been accounted for in $P_{15,\text{ctr}}$ and thus need to be subtracted from the result. The part that is new, is the one where only one of the loops becomes large and this limit will become part of the three-loop counterterm.
\item $P_{26}$: The double hard limit of this term is degenerate with $P_{24,\text{ctr}}$ and thus needs to be subtracted from the calculation. The single hard limit will lead to a new counterterm of the form $k^2 \times \text{finite}$.
\item $P_{35-I}$: The nature of this term is very similar to $P_{24}$, just extended by one loop of the stochastic kind. There are no degeneracies with lower order counterterms but it leads to a new two-loop counterterm of the form $k^2 \times \text{finite}$.
\item $P_{35-II}$ This term is a product of $P_{11}$, $P_{13}/P_{11}$ and $P_{15}/P_{11}$. The $k^2 P_{11}$ part of the $P_{15}$ contribution needs to be subtracted from the diagram. Finally, the counterterm is a combination of $c_s^2 k^2 \bar P_{15}$ and $c_s^2 k^2 \bar P_{15,\text{ctr}}$
\item $P_{44-I}$ is of the same nature as $P_{33-I}$ and $P_{22}$, \emph{i.e.}, the limits of high loop momenta lead to stochastic terms and the amplitudes of the divergencies are suppressed by $q^4$. 
\item $P_{44-II}$ The double hard limit with both ``ear" diagrams large is of the form $k^4 \sigma_d^4 \times \text{finite}$ and thus leads to a new counterterm.
\end{itemize}

In summary, we can conjecture for general $l$-loop diagrams that the diagram in which all the legs of the $F_{l+1}$ kernels are connected to another $F_{l+1}$, are stochastic terms and will not contribute to the leading order counterterm. All the other diagrams are dressings of diagrams that were encountered at lower orders with "ears", i.e. diagrams that are given by the pair of momenta $(\vec q,-\vec q)$ in the kernels. When one of them is hard, we need a new counterterm at leading order in derivatives and when several of them are hard, this term will be degenerate with a counterterm that has already been fixed. Thus their contribution has to be subtracted from the finite part of the loop calculation.

\section{Conclusions}

Our study leads to the following conclusions:
\begin{itemize}
\item Fitting the one-loop counterterm $\cs$ at $k=0.2 \ihMpc$ overestimates this coefficient by $50\%$ for lower wavenumbers and thus leads to too much suppression on large scales. While this regime is hard to extract from simulations, this might have important consequences for studies of primordial non-Gaussianity, where both the maximum wavenumber but also precision on large scales matter.
\item We find that the finite part of the two-loop calculation that is not degenerate with the leading order counterterm induces a $1\%$ correction in the power spectrum at $k=0.1\ihMpc$ for our $\Lambda$CDM cosmology at redshift $z=0$. The two-loop counterterm itself adds another $1\%$ correction in the same direction at $k=0.16\ihMpc$.
\item We have not explicitly calculated the three-loop correction, but based on a related study in Lagrangian space \cite{Baldauf:2015tla} and its extension to density fields \cite{Baldauf:2015tlb}, we estimate that the stochastic term will lead to a $1\%$ ($3\%$) correction at $k\approx 0.25\ihMpc$ ($k\approx 0.3\ihMpc$). Note that the perturbative EFT should describe the deterministic part
of the power spectrum, which by definition has less power than the
full non-linear power spectrum. Our EFT model fails by overpredicting
the power compared to simulations so the failure of the model really
happens at a slightly lower wavenumber.
Also, there does not appear to be much room for the explicit three loop calculation to improve on our results without explicitly modelling the stochastic term. 
\item We find that the two-loop EFT model with IR resummation can capture the scale dependence of the matter power spectrum up to $k\approx 0.3 \ihMpc$ at the $1\%$ level. Besides that, with the same EFT parameter and our assumption of a particular ($D^2$) time dependence of the counterterms, we are also able to explain the scale dependence of the momentum power spectrum.
\item The fact that the EFT model can match the power spectrum does not necessarily mean that the EFT is the right model for the density field. The model so far does not include the stochastic term, that is partially given by the virialized motions within haloes. We do expect the one halo term to play a role at the percent level around $k\approx 0.3 \ihMpc$ based on our recent study \cite{Baldauf:2015tlb}. The one halo term is just one contribution to the stochastic term.
\item At the sub-percent level, many numerical effects can affect the agreement between theory and simulations. On the one hand, one can use the perturbative calculations on very large scales to check linear growth in simulations. On the other hand, precise comparisons of simulations and theory on large scales might require the theory to be evaluated for the same seeds that were used to initialize the simulations as we discuss in App.~\ref{sec:propa}.
\item In summary, the picture at this point looks fairly consistent, but it is certainly too early to claim final success. 
Measurements of $\cs$ without cosmic variance on large scales as discussed in Appendix ~\ref{sec:propa} would certainly be the cleanest way to get the leading order EFT correction as well as its time dependence. Unfortunately there seem to be convergence issues in the $N$-body simulation. Besides that there might be several other effects that might warrant further study. For example,  the assumption of the EdS correspondence for the growth factor might impact our results at the percent level once the one-loop corrections become order unity corrections \cite{Takahashi:2008yk}.
\end{itemize}

\acknowledgements
The authors would like to thank Guido d'Amico, Mehrdad Mirbabayi, Roman Scoccimarro, Uro\v{s} Seljak, Leonardo Senatore, Marko Simonovi\'{c} and  Zvonimir Vlah for fruitful discussions.
T.B. is supported by the Institute for Advanced Study through a Corning Glass Works foundation fellowship.
L.M. was supported by the Swiss National Science Foundation throughout the initial phase of this project.
M.Z. is supported in part by the NSF grants  PHY-1213563 and AST-1409709.

\bibliography{twoloop}

\appendix

\section{Explicit Formulae}\label{app:expressions}
In this Appendix we write down the explicit formulae for the one- and two-loop contributions to set the normalization for the terms used in the main text. The corresponding diagrams are given in Fig.~\ref{fig:loopdiag}. For the one-loop terms we have
\beq
\begin{split}
P_{13}=&3 P(k)\int_{\vec q_1}F_3(\vec q_1,-\vec q_1,\vec k)P(q_1)\; ,\\
P_{22}=&2 \int_{\vec q_1}|F_2(\vec q_1,\vec k-\vec q_1)|^2P(q_1)P(|\vec k-\vec q_1|)\; ,
\end{split}
\eeq
where the recursion formulae for the SPT kernels can be found in the literature \cite{Bernardeau:2001qr}. The kernels employed here are the ones derived for an Einstein de-Sitter (EdS) Universe and are based on separability of spatial and temporal structure of the theory.
Note that all the kernels are symmetrized over their arguments. Their time dependence is governed by the $D^2$ scaling of the linear power spectrum with the growth factor $D$.
For the two-loop terms we have
\beq
\begin{split}
P_{15}=&15 P(k)\int_{\vec q_1}\int_{\vec q_2}F_5(\vec q_1,-\vec q_1,\vec q_2,-\vec q_2,\vec k)P(q_1)P(q_2)\; ,\\
P_{24}=&12 \int_{\vec q_1}\int_{\vec q_2}F_4(\vec q_1,-\vec q_1,\vec q_2,\vec k-\vec q_2)F_2(-\vec q_2,-\vec k+\vec q_2)P(q_1)P(q_2)P(|\vec k-\vec q_2|)\; ,\\
P_{33-I}=&6 \int_{\vec q_1}\int_{\vec q_2}|F_3(\vec q_1,\vec q_2,\vec k-\vec q_1-\vec q_2)|^2P(q_1)P(q_2)P(|\vec k-\vec q_1-\vec q_2|)\; ,\\
P_{33-II}=& P(k)\left[3\int_{\vec q_1}F_3(\vec q_1,-\vec q_1,\vec k)P(q_1)\right]^2=\frac{P^2_{13}(k)}{P(k)}\; .
\end{split}
\eeq
Note that we have not symmetrized the $P_{ij}$ terms for $i\neq j$, \emph{i.e.}, their contribution to the total equal time power spectrum will be $2 P_{ij}$.

\section{Cosmic variance and the propagator}\label{sec:propa}
\begin{figure}[t]
\centering 
\includegraphics[width=0.49\textwidth]{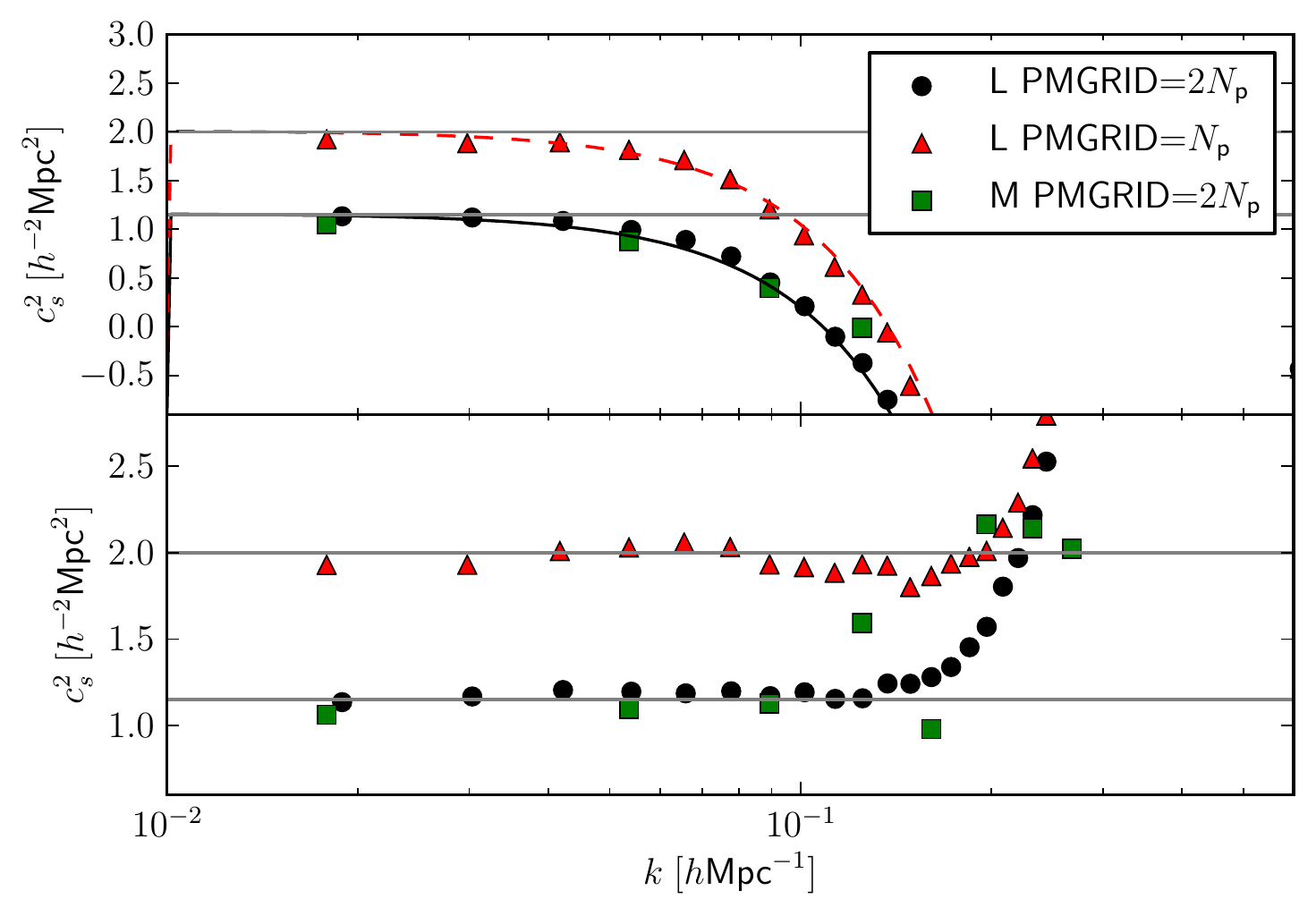}
\includegraphics[width=0.49\textwidth]{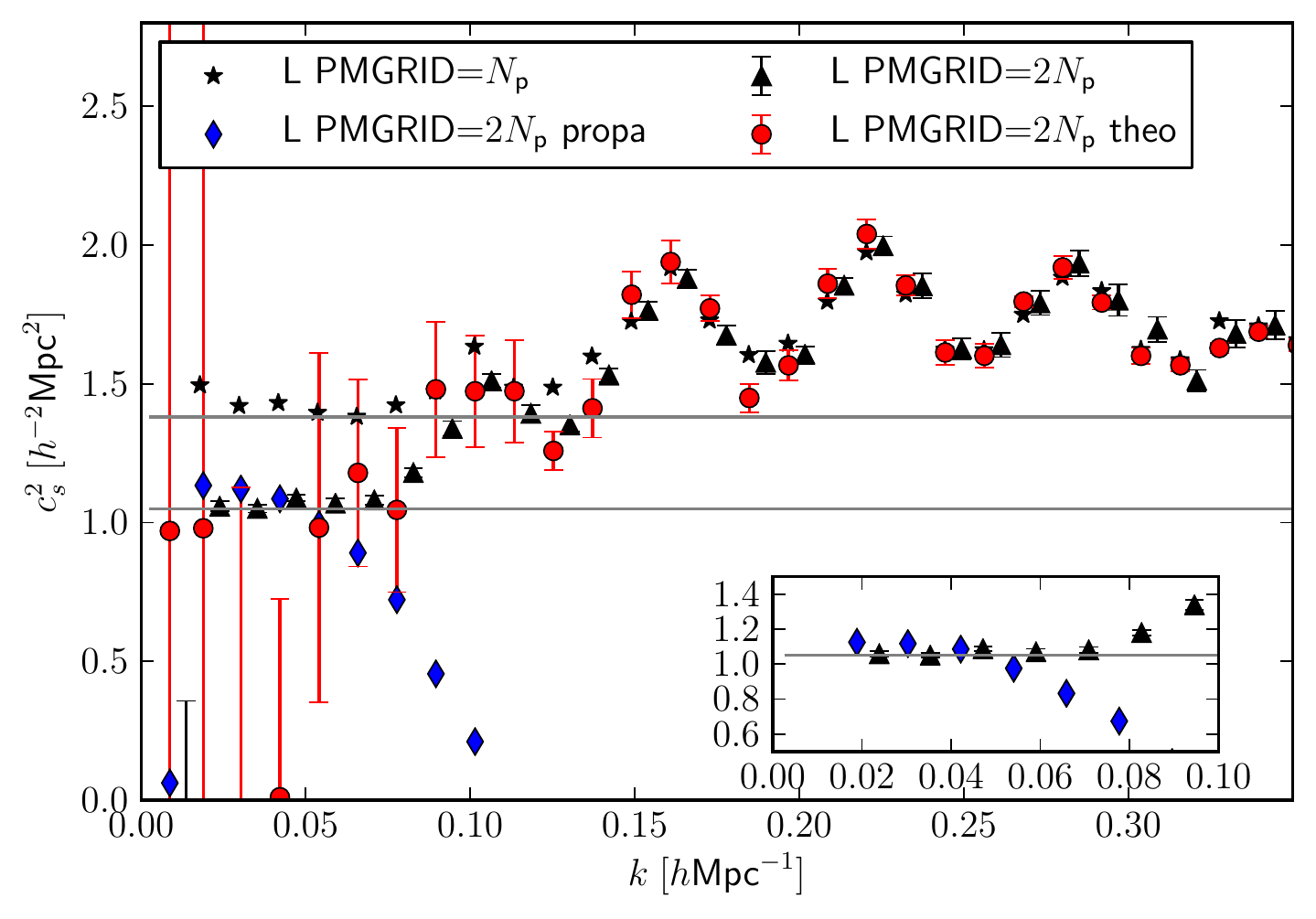}
\caption{\emph{Left panel: }Speed of sound extracted from propagator measurements in our simulations. We show measurements for two different choices of the Gadget PMGRID parameter $\text{PMGRID}=N_p$ and $\text{PMGRID}=2N_p$ for the L simulation as well as for the M simulation with $\text{PMGRID}=2N_p$. The upper panel shows the $k^2 P$ coefficient after the subtraction of the linear and one-loop corrections and the lower panel after subtraction of the regularized two-loop contribution $P_{15}^\text{reg}$ (which are shown as solid and dashed lines in the upper panel). \emph{Right panel: }Comparison of the $\cs$ estimator based on an analytical one-loop calculation and the one-loop calculation performed on the simulation IC grid. Once corrected for $3\times10^4$ growth factor normalization the low-$k$ limit of the estimator in the simulations indeed asymptotes to a constant.}
\label{fig:propa}
\end{figure}

The propagator \cite{Crocce:2005xy} measures the response of the density field to the initial conditions 
\beq
(2\pi)^3\ddir(\vec k-\vec k')G(k)=\la \frac{\partial \delta_\text{nl}(\vec k)}{\partial \delta_0(\vec k')}\ra \;.
\eeq
It can be estimated from the cross power spectrum of the non-linear and the linear field and the auto power spectrum of the linear field as
\beq
G(k)=\frac{P_{\text{nl},1}(k)}{P_{11}(k)}\; .
\eeq
The propagator is in principle the cleanest observable for the extraction of the speed of sound $\cs$ of the Eulerian EFT, since the latter is nothing but a modification of the low-$k$ limit of the propagator. At one-loop level we have $G=1+P_{13}/P_{11}-\cs k^2$ and at two loop level $G=1+P_{13}/P_{11}-\cs k^2+\bar P_{15}/P_{11}+\bar P_{15,\text{ctr}}/P_{11}$. Here, we again use the regularized $\bar P_{15}$, from which all the $k^2 P$ contributions have been removed. This is consistent with the previously introduced logic, where $\cs$ absorbs the difference after the one loop calculation has been subtracted from the data and effectively contains all the higher order and non-perturbative corrections as well as the mistake in the one-loop calculation. As argued before, this choice is somewhat arbitrary. This is no problem, as long as this choice is employed consistently.
RPT \cite{Crocce:2005xy} tries to resum the propagator in the high-$k$ limit, whereas here our goal will be to estimate the leading order corrections in the low-$k$ limit.

We are interested in the deviations of the propagator from unity and define an alternative estimator for $\cs$
\beq
\hat c_\text{s}=-\frac{P_{\text{nl},1}-P_{11}-P_{13}}{k^2 P_{11}} \;, \hspace{2cm} \hat c_\text{s}=-\frac{P_{\text{nl},1}-P_{11}-P_{13}-\bar P_{15}}{k^2 P_{11}} \;. 
\eeq
The SPT contributions in these expression are IR sensitive, since the cancellation of IR modes in $2P_{13}+P_{22}$ or its two-loop equivalent are not happening. When subtracting the one-loop SPT contribution from the propagator measured in simulations we address this issue by evaluating perturbation theory on the initial condition grid employed for the simulations using a technique similar to \cite{Roth:2011ru}. Thus, we are using the same IR modes that affect the non-linear dynamics in the simulation, thus directly addressing the IR sensitivity. The propagator based estimator projects out all the terms from the field that correlate with the linear field and thus provides an alternative and in some sense cleaner measurement of corrections that have the form of the leading order EFT counterterm, which could be masked by other contributions in the auto-power spectrum.

We show the measurements of $\cs$ based on the propagator in Fig.~\ref{fig:propa} at one- and two-loop level. The upper panel shows $\cs$ before the two-loop contribution has been subtracted. Once corrected for $10^{-4}$ level offsets in the linear growth factor,\footnote{We have independent evidence for such an error at this level from comparing power spectra for our fiducial parameter settings with a simulation with smaller timesteps.} that are probably related to the timestepping in \texttt{GADGET} and that lead to an upturn or downturn of the data points in the plot, we see that the data asymptote to a constant on large scales and then decay on smaller scales. The shape of this decay is however captured by the scale dependence of $\bar P_{15}$.
Another remarkable observation is that the measured value of $\cs$ depends very strongly on the PMGRID parameter in \texttt{GADGET}, leading to a shift of $\Delta \cs \approx 0.8 \hMpcsq$. We had seen a similar sensitivity already for the Lagrangian EFT coefficient of the displacement field in \cite{Baldauf:2015tla}. Based on this study, we are inclined to favor the results of the $\text{PMGRID}=2 N_p$ case. For another observation of this sensitivity in the power spectrum see \cite{Smith:2012uz}. Further evidence for the trustworthiness of this case comes from the fact that it agrees with the results from the higher resolution box, the M simulation.
In a second step we now remove the scale dependence of $\bar P_{15}$ and see in the lower panel, that the estimated $\cs$ is flat up to $k\approx 0.2\ihMpc$. There is clearly more need for convergence studies of the propagator and we certainly do not want to overinterpret a result that is so sensitive on numerical parameters of simulations. We conclude however, that there is evidence for a non-zero speed of sound correction after one loop SPT has been subtracted. The inferred value $\cs=1.15 \ihMpc$ roughly agrees with the value employed for the equal time correlators in the main text.
Had we subtracted the explicit low-$k$ limit of $P_{15}$ (not $\bar P_{15}$, i.e. before regularization), the estimated $\cs$ would have changed by $\Delta \cs=-2.58\ihMpc$ and thus yielded a non-zero, negative $\cs$. The explicit corrections from three loops are even higher \cite{Blas:2013aba}.

In the right panel of Fig.~\ref{fig:propa} we also compare the grid based calculation for the $\cs$ constraint from the auto power spectrum with the analytical calculation. For the latter, we saw in the main text in Fig.~\ref{fig:fitcs} that the data points for $\hat c_\text{s}^2$ at $k<0.05\ihMpc$ are systematically low. This problem vanishes once the theory is calculated on the simulation grid. We see that the $\cs$ estimator now asymptotes to a constant horizontal line on large scales, as we would expect it to based on the scale dependence of the two-loop corrections shown in Fig.~\ref{fig:csscaldep}. Again, the value of this asymptotic constant depends strongly on the PMGRID parameter choice, now leading to a $\Delta \cs = 0.3 \hMpcsq$ difference between the two cases. Note however that they agree at higher wavenumbers. Thus, to the extend that our ansatz is trustworthy, a model what matches at these scales would prefer the $\text{PMGRID}=2N_p$ case at lower wavenumbers.

\begin{figure}
\includegraphics[width=0.49\textwidth]{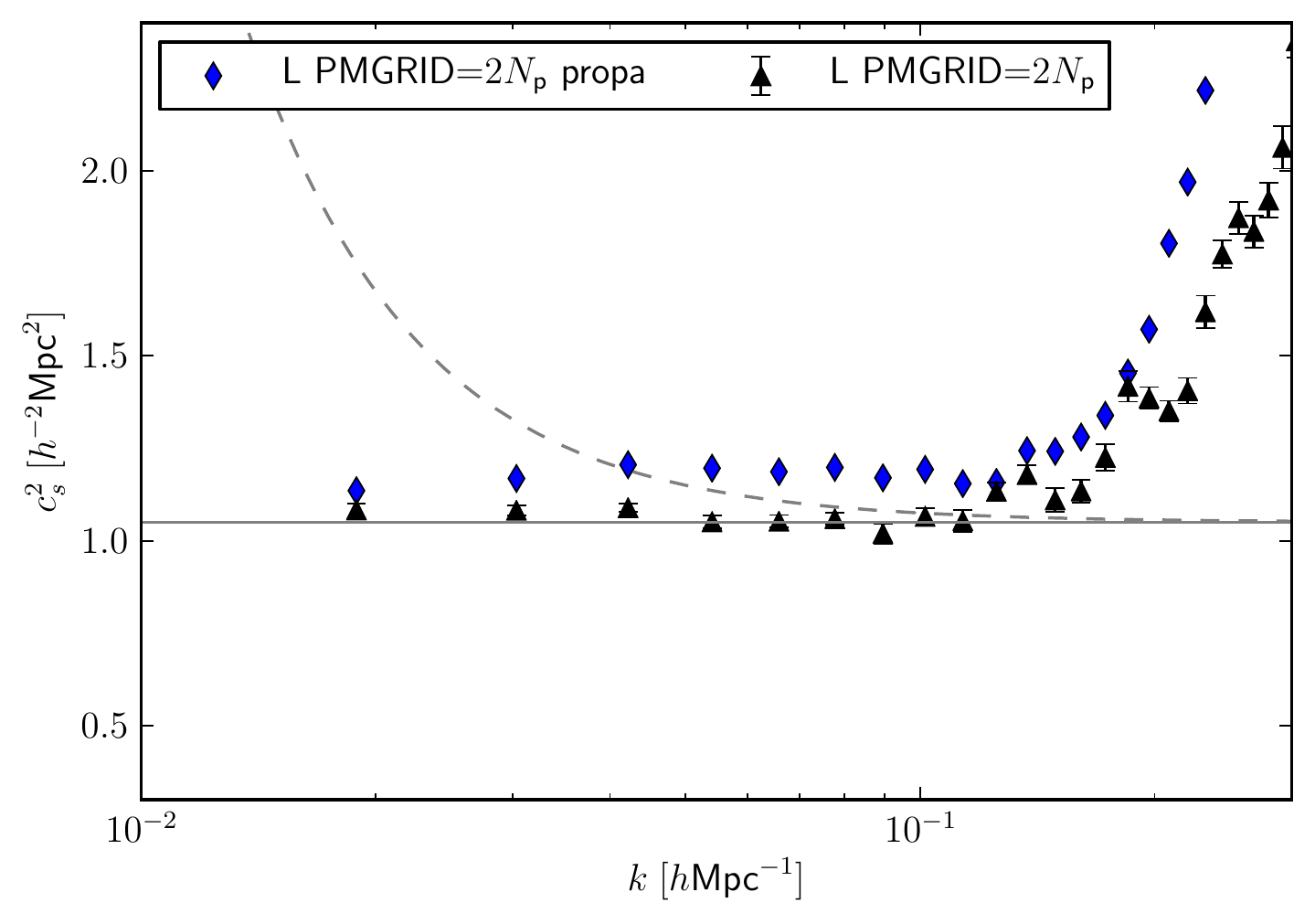}
\caption{Low-$k$ measurements of $\cs$ from the final field and the propagator after the two-loop terms have been corrected for. The dashed line shows the effect of a relative error of the linear growth factor of $2\times 10^{-4}$ that both statistics have been corrected for.}
\label{fig:lowk}
\end{figure}

There is also a slight disagreement between the propagator and power spectrum estimates for the favored PMGRID$=2N_p$ case. The power spectrum method of this case would indicate a $\cs=1.05 \hMpcsq$. In Fig.~\ref{fig:lowk} we show both the propagator and the power spectrum estimator after the finite two loop terms have been subtracted out. Except for a $\Delta c_s^2\approx 0.1 \hMpcsq$ offset both estimators are flat and consistent up to $k\approx 0.15 \ihMpc$, where higher order terms, for instance the two loop counterterms, start to matter.

\section{Limits of the two loop terms}\label{app:limits}
\begin{table}[t]
\begin{tabular}{l|ccc|ccc}
& \multicolumn{3}{c|}{single-hard}& \multicolumn{3}{c}{double-hard}\\
& Eq.&$x(n)$&$x(-3/2)$& Eq.&$x(n)$&$x(-3/2)$\\
\hline
$P_{15}$ &\eqref{eq:p15singlehard}&n+1&-1/2 &\eqref{eq:p15doublehard}&2n+4&1\\
$P_{24}$&\eqref{eq:p24singlehard}  &n+1&-1/2& \eqref{eq:p24Idoublehard}&3n+2&-5/2\\
& \eqref{eq:p24singlehard2} &2n-1&-4&& &\\
$P_{33-I}$&\eqref{eq:p33Isinglehard}&2n-1&-4&\eqref{eq:p33Idoublehard}&3n+2&-5/2\\
$P_{33-II}$&\eqref{eq:p33IIsinglehard}&n+1&-1/2&\eqref{eq:p33IIdoublehard}&2(n+1)&-2\\
\end{tabular}
\caption{Table of the two loop limits, references to the equations where they are discussed, the power of the cutoff dependence $\Lambda^x$ for a power law power spectrum $P(k)\propto k^n$ with general power law slope $n$ and for $n=-3/2$. For the single hard limit the slope gives the power of the hard integral ignoring the remaining finite integral, while for the double hard integrals we consider both momenta in the hard integrals to be of the same order. The choice $n=-3/2$ is motivated by the slope of our $\Lambda$CDM power spectrum at $k\approx 0.1 \ihMpc$.}
\label{tab:twolooplimits}
\end{table}
In the main text, we have concentrated our discussion on the terms that we consider relevant for the leading UV sensitivity and the corresponding counterterms. Let us, for the sake of completeness, discuss the remaining hard limits in this appendix. An overview of all the single- and double-hard limits of the two loop calculation is given in Tab.~\ref{tab:twolooplimits}. In this table we also give the power of the cutoff dependence of the remaining integrals if the initial power spectrum is of power law form $P(k)\propto k^n$. We evaluate the cutoff dependence for $n=-3/2$, the slope of our power spectrum at $k=0.1\ihMpc$. For the single-hard limits we immediately see that the terms that we found to dominate the shell behaviour have the most shallow decay in the UV, and are thus the most sensitive to the change of the power spectrum at high wavenumbers. For the double hard limits, the limit of $P_{15}$ is still growing for $n=-3/2$ but turns around at for $n=-2$ at $k\approx 0.3$, so it will still converge based on the high-$k$ slope of our initial power spectrum. Yet, it is immediately clear why this integral should be absorbed into the counterterm. The subleading $k^4 P$ UV-sensitivity of $P_{15}$ (not mentioned in the table but below in Eq.~\ref{eq:p15doublehard}) scales as $2n+2$, i.e., as $\Lambda^{-1}$ for $n=-3/2$ and should thus be the next term considered as a counterterm, after the single hard limits. It will change the coefficient of the $c_s^4 k^4 P_{11}$ counterterm.

First, we consider the limit of $P_{33-I}$ for $q_1\to \infty$ while $q_2$ remains finite
\beq
\begin{split}
P_{33-I}^{q_1\to\infty}=& \int_{\vq_1}\int_{\vq_2} \frac{P^2_{11}(q_1)}{r_1^4}P_{11}(q_2) \Biggl[\frac{5565 r_2^9-11465 r_2^7+26409 r_2^5+22285 r_2^3-510 r_2}{14288400 r_2^5}\\
&-\frac{15 \left(r_2^2-1\right)^3 \left(7 r_2^2+2\right) \left(53 r_2^2+17\right) }{14288400 r_2^5} \log\left(\frac{1+r_2}{1-r_2}\right)\Biggr] \;.
\label{eq:p33Isinglehard}
   \end{split}
\eeq
The amplitude is given by $\int P_{11}^2(q)/q^4$, which is quickly convergent in the UV and does thus not contribute to a significant UV sensitivity. For the double-hard limit of the same integral we have
\beq
\begin{split}
P_{33-I}^{q_1,q_2\to\infty}=& k^4\int_{\vec q_1}\int_{\vec q_2} K_{33-I}(\vec q_1,\vec q_2) P_{11}(q_1)P_{11}(q_2)P(|\vec q_1+\vec q_2|)\; ,
\label{eq:p33Idoublehard}
   \end{split}
\eeq
where $K_{33-I}$ can be parametrized in terms of the magnitude of the momenta and their cosine $\mu_{12}=\vec q_1\cdot \vec q_2 /q_1q_2$ as
\beq
K_{33-I}=\frac{\left(  \mu_{12}^2-1\right)^2 \left(q_1q_2 \left(392   \mu_{12}^3 +634   \mu_{12} \right)+\left(392
  \mu_{12}^2+121\right) q_1^2+\left(392   \mu_{12}^2+121\right) q_2^2\right)}{19845 \left(q_1^2+2   \mu_{12} q_1
q_2+q_2^2\right)^3}\; .
\eeq
The term thus scales as $k^4$ in the $k\to 0$ limit, with a quickly convergent amplitude $ \int \int P_{11}^3/q^4$.

For the double-hard limit of $P_{24}$ we have
\beq
\begin{split}
P_{24}^{q_1,q_2\to\infty}=& k^4\int_{\vec q_1}\int_{\vec q_2} K_{24}(\vec q_1,\vec q_2) P_{11}(q_1)P_{11}(q_2)P_{11}(|\vec q_1+\vec q_2|)\;,
\label{eq:p24Idoublehard}
   \end{split}
\eeq
where $K_{24}$ is given by
\beq
\begin{split}
K_{24}=&\frac{4 \left(\mu _{12}^2-1\right)^2}{33957 q_2^2  \bigl(2 \left(1-2 \mu _{12}^2\right) q_2^2
   q_1^2+q_1^4+q_2^4\bigr)^2} \Bigl[\left(1380 \mu _{12}^2-43\right) q_1^6+\left(-2544 \mu _{12}^4+208 \mu
   _{12}^2+243\right) q_2^2 q_1^4\\
   &+\left(-1008 \mu _{12}^4+568 \mu _{12}^2+615\right) q_2^4 q_1^2+7 \left(36 \mu_{12}^2+47\right) q_2^6\Bigr] .
\end{split}
\eeq
The term thus scales as $k^4$ in the $k\to 0$ limit, with a quickly convergent amplitude.

In the main text we considered only one of the single-hard limits of $P_{24}$, the one where the closed loop or ``ear" diagram becomes large. The other limit, where the loop containing two power spectra becomes large is given by
\beq
P_{24}^{q_2\to \infty}=k^4\int_{\vec q_2}\frac{P_{11}^2(q_2)}{q_2^4}\int_{\vec q_1}K_{24}(r_1) P_{11}(q_1)\; ,
\label{eq:p24singlehard2}
\eeq
with a kernel
\beq
K_{24}=\frac{28620 r_1^9-90858 r_1^7+470848 r_1^5-247326 r_1^3+4644 r_1-3 \left(r_1^2-1\right){}^3
   \left(4770 r_1^4-2423 r_1^2-774\right) \log
   \left(\frac{r_1+1}{r_1-1}\right)}{16299360 r_1^5}\; .
\eeq
As for $P_{33-I}^{q_1\to \infty}$, the amplitude is given by $\int P_{11}^2(q)/q^4$.

As we had hinted at in the main text, if both loop momenta in $P_{15}$ go to infinity, we end up with  an analytic function in $k$
\beq
P_{15}^{q_1,q_2\to\infty} =-k^2 P_{11}(k)\int_{\vec q_1}\int_{\vec q_2} K_{15,2}(q_1,q_2)P_{11}(q_1)P_{11}(q_2) +k^4 P_{11}(k) \int_{\vec q_1}\int_{\vec q_2} K_{15,4}(q_1,q_2)P_{11}(q_1)P_{11}(q_2)\; .
\label{eq:p15doublehard}
\eeq

We rederive the double hard limit of $P_{15}$ and note that the expression agrees with what was found for the $k^2P_{11}$ part in \cite{Bernardeau:2012ux,Blas:2013bpa}\\
\beq
\begin{split}
K_{15,2}(q_1,q_2)=&-\frac{2 q_1 q_2 \left(5760 q_1^{10}+19365 q_1^8 q_2^2-114653 q_1^6 q_2^4-114653
   q_1^4 q_2^6+19365 q_1^2 q_2^8+5760 q_2^{10}\right)}{13759200 q_1^7 q_2^7}\\
   &+\frac{15 \left(q_1^2-q_2^2\right)^4
   \left(384 q_1^4+2699 q_1^2 q_2^2+384 q_2^4\right) \log
   \left(\frac{q_1+q_2}{q_1-q_2}\right)}{13759200 q_1^7 q_2^7}\; .
\end{split}
\eeq
Note that the limit of the above expression, where one of the momenta is smaller than the other one $q_2 \ll q_1$, turns out to be the same as the limit of Eq.~\eqref{eq:p15hardsoft}, in which we first take $q_1\gg k,q_2$ and only afterwards $q_2 \gg k$. We also obtain the expression for the $k^4 P_{11}$ part to estimate the size of the subleading UV-sensitivity
\beq
\begin{split}
K_{15,4}(q_1,q_2)=&\frac{1}{59329670400 q_1^9
   q_2^9}\Bigl(7074480 q_1^{13} q_2+75154450 q_1^{11} q_2^3-41165824 q_1^9 q_2^5+317583580
   q_1^7 q_2^7\\
   &-41165824 q_1^5 q_2^9+75154450 q_1^3 q_2^{11}+7074480 q_1 q_2^{13}\Bigr)\\
   &-\frac{105 \left(q_1^2-q_2^2\right)^2}{59329670400 q_1^9 q_2^9} \log  \left(\frac{q_1+q_2}{q_1-q_2}\right) 
   \Bigl(33688 q_1^{10}+414025 q_1^8 q_2^2+476047 q_1^6\\
   &  q_2^4+476047 q_1^4 q_2^6+414025 q_1^2 q_2^8+33688 q_2^{10} \Bigr)\; .
\end{split}
\eeq
\end{document}